\documentclass[aps,pra,twocolumn,showkeys,preprintnumbers,superscriptaddress]{revtex4-1}
\usepackage{lettrine}

\usepackage[american]{babel}
\usepackage{graphicx}
\usepackage{amsmath}
\usepackage{amssymb}
\usepackage{amstext}
\usepackage{amsthm}
\usepackage{latexsym}
\usepackage{array}
\usepackage{color}
\usepackage{float}
\usepackage{microtype}
 
\usepackage{multirow}
\usepackage{dcolumn}
\usepackage[caption=false]{subfig}
\definecolor{url}{RGB}{0,20,160}
\usepackage[colorlinks=true,linkcolor=blue,citecolor=blue,urlcolor=url]{hyperref}
\usepackage[usenames,dvipsnames,svgnaes,table]{xcolor}
\usepackage[usenames,dvipsnames,svgnaes,table]{xcolor}
\definecolor{col1}{HTML}{65717D}
\definecolor{col2}{HTML}{1C5CFF}
\definecolor{col3}{HTML}{FF1654}
\definecolor{col4}{HTML}{00A300}
\definecolor{col5}{HTML}{D500DE}

\definecolor{col1h}{HTML}{DB5A42} %orange
\definecolor{col2h}{HTML}{19A0D1} %blue
\definecolor{col3h}{HTML}{57A773} %green
\definecolor{col4h}{HTML}{CC5981} %onion color
\definecolor{col5h}{HTML}{4F6D16} %onion color
\definecolor{col6h}{HTML}{028090} %onion color

\begin{document}
\title{Achieving ultra-high power factor in Sb$_{2}$Te$_{2}$Se via valence band convergence}   
\author{Mohammad Rafiee Diznab$^{\dagger}$}
\affiliation{Department of Physics, University of Tehran, Tehran 14395-547, Iran} 
\altaffiliation{Contributed equally to this work}
\author{Iraj Maleki Shahrivar$^{\dagger}$}
\affiliation{Department of Physics, University of Tehran, Tehran 14395-547, Iran}
\altaffiliation{Contributed equally to this work}
\author{S. Mehdi Vaez Allae}
\affiliation{Department of Physics, University of Tehran, Tehran 14395-547, Iran} 
\author{Yi Xia}
\affiliation{Department of Materials Science and Engineering, Northwestern University,	Evanston, Illinois 60208, USA} 
\author{S. Shahab Naghavi}
\email{s_naghavi@sbu.ac.ir}
\affiliation{Department of Physical and Computational Chemistry, Shahid Beheshti University, G.C., Evin, 1983963113 Tehran, Iran}
%%%%%%%%%%%%%%%%%%%%%%%%%%%%%%%%%%%%%%%%%%%%%%%%%%%%%%%%%%%%%%%%%%%%%%%%%%%%%%%%%
%\date{\today}
%%%%%%%%%%%%%%%%%%%%%%%%%%%%%%%%%%%%%%%%%%%%%%%%%%%%%%%%%%%%%%%%%%%%%%%%%%%%%%%%% 
\begin{abstract}
An efficient approach to improve the thermoelectric performance of materials is to converge their electronic bands, which is known as band engineering. In this regard, lots of effort have been made to further improve the thermoelectric efficiency of bulk and exfoliated monolayers of Bi$_{2}$Te$_{3}$ and Sb$_{2}$Te$_{3}$.  However, ultra-high band degeneracy and thus significant improvement of power factor have not been yet realized in these materials. Using first-principles methods, we demonstrate that the valley degeneracy of Bi$_{2}$Te$_{3}$ and Sb$_{2}$Te$_{3}$ can be largely improved upon substitution of the middle layer Te atoms with the more electronegative S or Se atoms. Our detailed analysis reveals that in this family of materials two out of four possible valence band valleys merely depend on the electronegativity of the middle layer chalcogen atoms, which makes the independent modulation of the valleys' position feasible. As such, band alignment of Bi$_{2}$Te$_{3}$ and Sb$_{2}$Te$_{3}$ largely improves upon substitution of the middle layer Te atoms with more electronegative, yet chemically similar, S and Se ones. A superior valence band alignment is attained in Sb$_{2}$Te$_{2}$Se monolayers where the three out of four possible valleys are well-aligned, resulting in a giant band degeneracy of 18 that holds the record among all thermoelectric materials.
  
\end{abstract} 
%%%%%%%%%%%%%%%%%%%%%%%%%%%%%%%%%%%%%%%%%%%%%%%%%%%%%%%%%%%%%%%%%%%%%%%%%%%%%%%%%
\pacs{xxxxx}
\maketitle
%%%%%%%%%%%%%%%%%%%%%%%%%%%%%%%%%%%%%%%%%%%%%%%%%%%%%%%%%%%%%%%%%%%%%%%%%%%%%%%%%
\noindent \textbf{\large Introduction} \\[0.063em]
\lettrine[lines=2,loversize=0.18]{\color{col1}\textbf{T}}{}he electro-thermal energy conversion is an environmentally friendly solution to the global energy crisis, yet the efficiency of energy conversion should be significantly improved for practical applications of thermoelectric (TE) devices~\cite{goldsmid2010introduction}. The efficiency of TE materials is determined by the thermoelectric figure of merit $zT=S^{2} \sigma/(\kappa_{e}+\kappa_{\rm L})$, where $\sigma$ is the electrical conductivity, $S$ is the Seebeck coefficient or thermopower, $T$ is the absolute temperature, $\kappa_{\rm e}$ and $\kappa_{\rm L}$ are electronic and lattice contributions to the thermal conductivity, respectively. Achieving an optimal $zT$ value is highly nontrivial, which requires dealing with conflicting parameters, since $S$ and $\sigma$ are inversely related, and $\kappa_{\rm e}$ is proportional to $\sigma$ as implied by the Wiedemann-Franz law ($\kappa_{\rm e} =  L\sigma T$). 

%Currently, there seems to be a two-fold solution to address the efficiency issue: 
To improve TE performance, various strategies have been proposed, which can be majorly categorized into either i) Phonon engineering \cite{tang2010holey,majumdar2004thermoelectricity,poudel2008high,dresselhaus1999low} or ii) Band engineering \cite{liu2012convergence,tang2015convergence,zhang2012heavy,witkoske2017thermoelectric,pei2011convergence,pei2012band,he2017bi2pdo4,he2019designing}. Thanks to the relatively decoupled nature of lattice thermal transport from the electronic properties, phonon engineering has been frequently utilized to suppress $\kappa_{L}$ through creating nanostructures, resulting in a direct boost in $zT$. On the other hand, band engineering can be typically realized via converging separate energy pockets near the Fermi level (valley degeneracy), thus enhancing the power factor ($PF$), which is defined as $PF=S^{2}\sigma$. It should be noted that, increasing $S$ only, for example, through increasing the electron effective mass ($m^{*}$), may not ultimately lead to $PF$ enhancement, because it is also detrimental to $\sigma$. Therefore, a better scenario is a combination of multiple conducting channels composed of degenerate valleys where every single valley provides a low $m^{*}_{\rm band}$ and thus high electron mobility~\cite{goldsmid2013thermoelectric,pei2011convergence}. Since the overall density-of-states effective mass is proportional to the orbital degeneracy, $m^{*}$=$N_{\rm v}^{2/3}m_{b}^{*}$, increasing the number of degenerate valleys simultaneously enhances the Seebeck coefficient, another advantage to attain high $PF$.

Accordingly, TE materials with intrinsically high $N_{\rm v}$ and low $m_{b}^{*}$ have long been under intensive study, such as PbTe, Bi$_{2}$Te$_{3}$, Sb$_{2}$Te$_{3}$, Zintl to name a few. Particularly, Bismuth Telluride and its alloys have attracted considerable amount of attention \cite{wright1958thermoelectric,goldsmid2013thermoelectric,kim2017high,yan2010experimental} as a practical realization of TE devices. Their excellent TE performance stems from (i) the six-fold degeneracy of the valence-band maximum in the bulk Bi$_{2}$Te$_{3}$~\cite{shi2015connecting} and (ii) the intrinsically low lattice thermal conductivity of about $\approx$1.5 $Wm^{-1}K^{-1}$\cite{goldsmid1956thermal}, resulting in an excellent $zT$ $\approx$1 at T\,$\approx\!320\,K$~\cite{tritt2006thermoelectric}. In addition, it has been demonstrated by Dresselhaus \textit{et. al.}~\cite{dresselhaus1999low} that the emergence of quantum confinement effect by reducing the dimensionality could further improve $zT$. This strategy has been successfully applied to Bi$_{2}$Te$_{3}$ nanowires, where a 13\% enhancement in $zT$ has been achieved compared to the $n$-type Bi$_{2}$Te$_{2.7}$Se$_{0.3}$ bulk counterpart~\cite{zhang2011rational}. Theoretically, exfoliated monolayers of Bi$_{2}$Te$_{3}$ has been predicted to attain an ultrahigh $zT$ of $\approx$2.7 at $700\, K$ \cite{Sharma_2016}.

The present work aims to improve TE efficiency of Bi$_{2}$Te$_{3}$ and Sb$_{2}$Te$_{3}$ monolayers, and to suggest design strategies for high-performance TE materials. We show that substitution of the middle layer Te atoms with S and Se improves the valence band alignment in both Bi$_{2}$Te$_{2}$S and Bi$_{2}$Te$_{2}$Se monolayers. Particularly we attain an outstanding valence band alignment with $N_{\rm v}$ of 18 in Sb$_{2}$Te$_{3}$Se monolayers, which is one of the highest reported $N_{\rm v}$ values to the best of our knowledge. Meanwhile, all the studied materials largely benefit from their intrinsically low $\kappa_{\rm L}$ due to the strong anharmonicity and heavy average atomic masses~\cite{slack1973nonmetallic,lindsay2013first,naghavi2018pd2se3,peng2016first}. Both of these two factors lead to a significant improvement in $zT$. Therefore, we suggest that solid solutions of Sb$_{2}$Te$_{3-x}$Se$_{x}$, Bi$_{2}$Te$_{3-x}$S$_{x}$ and Bi$_{2}$Te$_{3-x}$Se$_{x}$ with $x$=1, are more advantageous than  their parent compounds for TE applications. It is also worth noting that all the studied compounds in the present work are thermodynamically stable in their bulk phase \cite{ceder2010materials,ong2008li,jain2011formation} and have been predicted to be exfoliable with dynamically stable monolayers \cite{mounet2018two}. The bulk Sb$_{2}$Te$_{2}$Se, Bi$_{2}$Te$_{2}$S and Bi$_{2}$Te$_{2}$Se have been experimentally synthesized~\cite{kanagaraj2019structural}.

%%%%%%%%%%%%%%%%%%%%%%%%%%%%%%%%%%%%%%%%%%%%%%%%%%%%%%%%%%%%%%%%%%%%%%%%%%%%%%%%%
\vskip 0.7cm
\noindent \textbf{\large Computational Methods} \\ [0.5em]
In this study, we performed theoretical calculations using the Density Functional Theory (DFT) as implemented in the Vienna ab-initio Simulation Package (\texttt{VASP}) \cite{kresse1993ab,kresse1996efficiency}. The projector augmented wave (PAW) pseudo-potentials~\cite{blochl1994projector,kresse1999ultrasoft}, plane wave basis set, and Perdew-Burke-Ernzerhof (PBE) exchange-correlation functionals were used \cite{perdew1996generalized} throughout the calculations. The cut-off energy was set to 500\,eV and a $24\times24\times1$ $k$-mesh sampling was used to ensure tight energy convergence. All structures were fully relaxed with respect to lattices and positions until the forces on each atom become less than 0.1\,meV$\cdot$\AA$^{-1}$. Due to the importance of Spin Orbit Coupling (SOC), we also considered this correction when calculating the electronic structure. Electronic transport properties, namely $\kappa_{e}$, $\sigma$ and $S$, were calculated by solving the Boltzmann Transport Equation (BTE) within the constant relaxation time approximation as implemented in \texttt{BoltzTraP} \cite{madsen2006boltztrap}. Therein, the Brillouin zone (BZ) was sampled using a dense $k$-grid of 48$\times$48$\times$1 to ensure accurate interpolation of the Kohn-Sham eigenvalues for \texttt{BoltzTraP}. To obtain the second order Inter-atomic Force Constants (IFCs), we used the finite displacement method as implemented in \texttt{Phonopy}~\cite{togo2008first} code using a $4\times4\times1$ supercell with a $k$-mesh of $3\times3\times1$. Third order IFCs were calculated by the \texttt{ShengBTE} code \cite{li2014shengbte} using  $3\times3\times1$ supercells, with the corresponding interaction cut-off being set to the third nearest neighboring shell and $4\times4\times1$ $k$-mesh sampling of the Brillouin zone.

%%%%%%%%%%%%%%%%%%%%%%%%%%%%%%%%%%%%%%%%%%%%%%%%%%%%%%%%%%%%%%%%%%%%%%%%%%%%%%%%%
\vskip 0.5cm
\noindent \textbf{\large Results and Discussion} \\ [0.5em]
%%%%%%%%%%%%%%%%%%%%%%%%%%%%%%%%%%%%%%%%%%%%%%%%%%%%%%%%%%%%%%%%%%%%%%%%%%%%%%%%%
\noindent{\textbf{Crystal Structure.}}
\begin{figure}[htp!]
	\centering
	\includegraphics[width=1.0\linewidth]{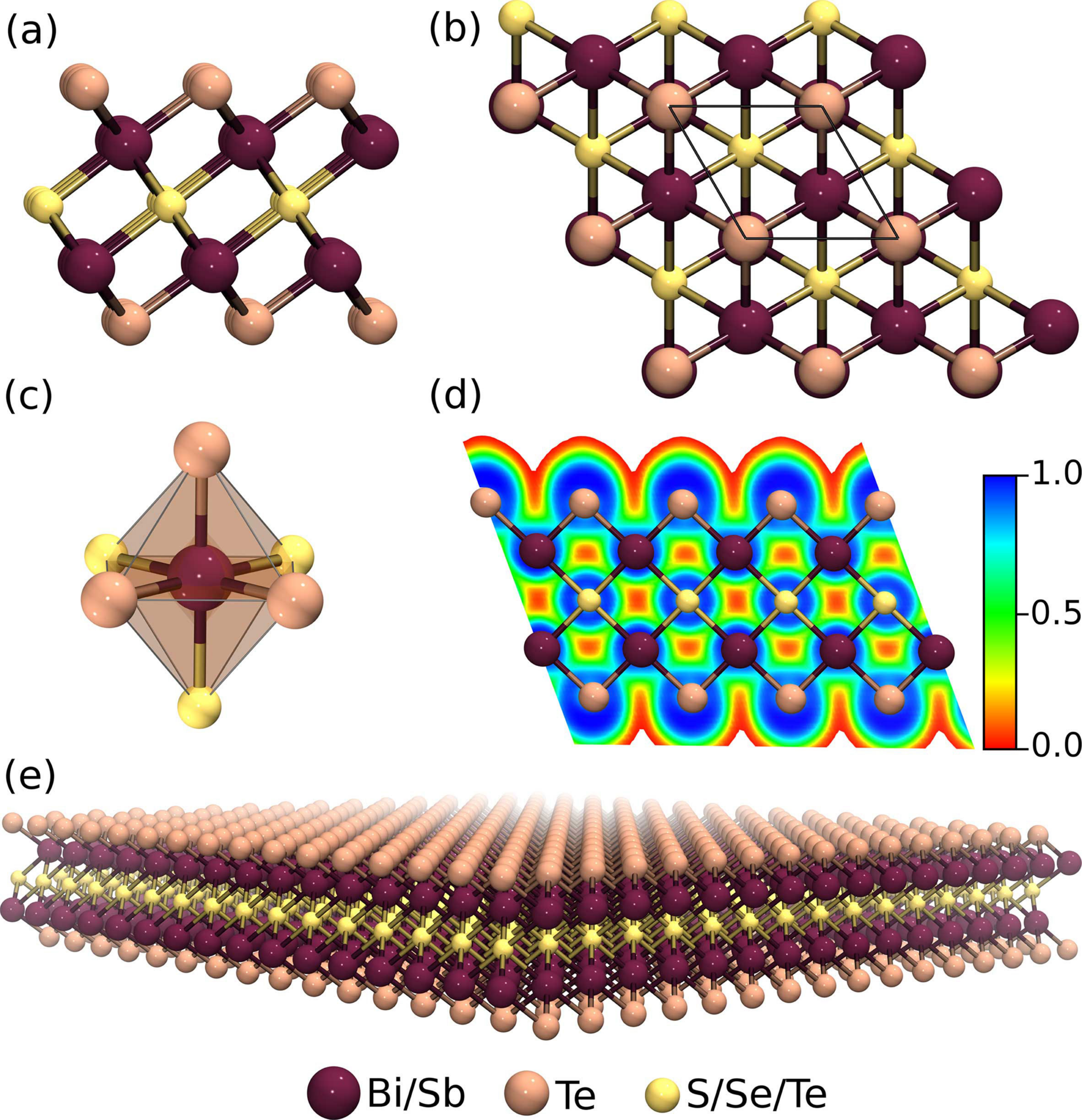}
	\caption{(a) Side and (b) top view of a quintuple layer ($QL$) of Bi$_{2}$Te$_{3}$ or equivalent materials. Middle layer Tellurium atom is different in its nature due to different coordination, as indicated in (c). The crystal structure consists of neighboring octahedra. (d) Electronic Localization Function (ELF) of the studied materials. Regions where the ELF is zero (one) correspond to no (full) localization. (e) Extended view of  the $M_{2}$Te$_{2}X$ (M=Bi, Sb; X=S, Se, Te) family of monolayers in which the middle layer is represented by yellow spheres.}
\label{FIG:Structure}
\end{figure}
Sb$_{2}$Te$_{3}$ and Bi$_{2}$Te$_{3}$ have the space group $R3\bar{m}$ with a rhombohedral crystal structure that belongs to the hexagonal crystal family. The bulk structure consists of quintuple layers, held together by the weak van der Waals (vdW) interaction \cite{luo2012first,bjorkman2012van}. According to existing experiments~\cite{mounet2018two,ambrosi2018exfoliation}, monolayers of Bi$_{2}$Te$_{3}$ and Sb$_{2}$Te$_{3}$ can be exfoliated from their bulk phase at an energetic cost of 25\,meV/{\AA}$^{2}$, which is comparable to the cost of peeling off a single layer of graphene from graphite. Therefore both compounds can be  categorized as experimentally exfoliable materials~\cite{mounet2018two}.

\begin{figure*}[htp!]
	\centering
	\includegraphics[width=1.0\linewidth]{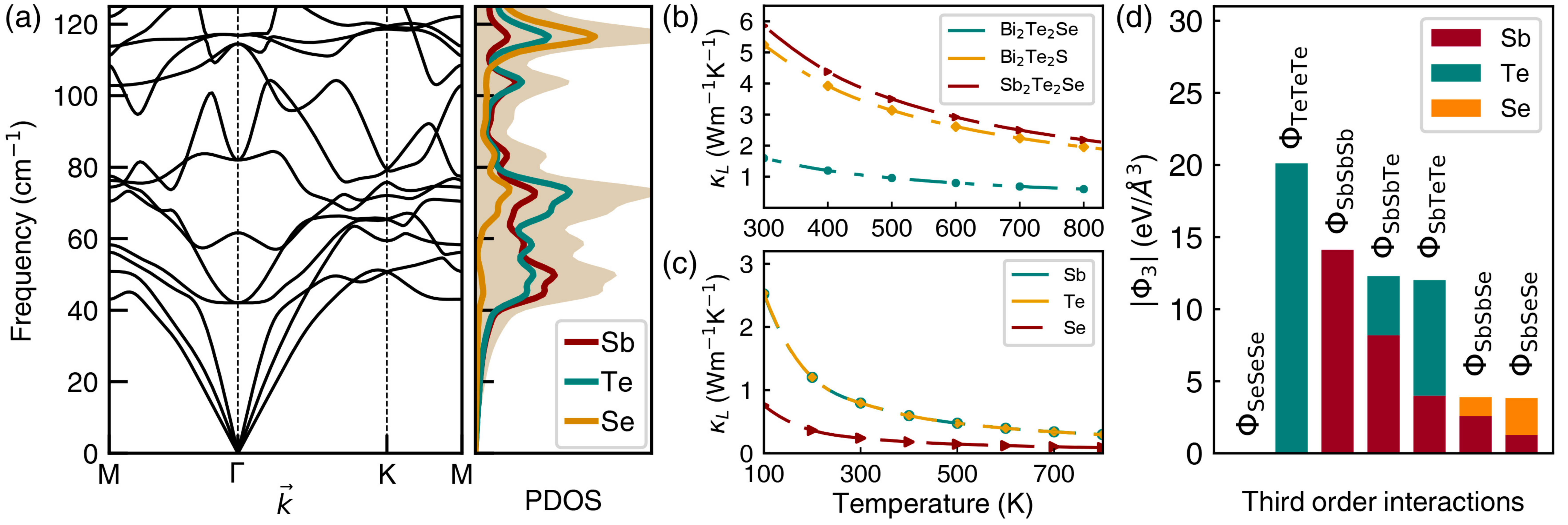}
	\caption{
          Phonon transport properties: (a) Phonon dispersion and the density of states (DOS) of Sb$_{2}$Te$_{2}$Se. Note the avoided crossings between optical and acoustic modes along the $\Gamma$-$M$ and $\Gamma$-$K$ directions. The partial DOS indicates the rather similar contributions from Sb and Te atoms to the total DOS, as well as the enhanced contribution from Se atoms in the middle layer with increased frequency. (b) Lattice thermal conductivities of the studied materials as a function of temperature from 300 to 800 K. Due to the higher average mass of Bi$_{2}$Te$_{2}$Se, its $\kappa_{L}$ is significantly smaller than that of Sb$_{2}$Te$_{2}$Se. (c) Atom-decomposed $\kappa_{L}$ of Sb$_{2}$Te$_{2}$Se as a function of temperature. The decomposition is according to the amplitudes of the atomic eigendisplacements, showing that Sb and Te have similar contributions to $\kappa_{L}$ across the whole temperature range, while the contribution from Se is significantly smaller. (d) Frobenius norm of the calculated third-order inter-atomic force constants, indicating the magnitude of anharmonicity. The larger magnitude of $\Phi$ associated with Te atoms is attributed to the lone-pair electrons.}
\label{FIG:Phonon}
\end{figure*}

A closer look over the crystal structure of Bi$_{2}$Te$_{3}$ (likewise Sb$_{2}$Te$_{3}$) reveals that there are two unique kinds of Te atoms in each monolayer, and thus can be labeled as Bi$_{2}$[Te$_{(1)}$]$_{2}$[Te$_{(2)}$].
In the Bi$_{2}$Te$_{3}$ monolayers, the number of valance electrons responsible for bonding reaches twenty-four per unit-cell, with four from each of the outer layer Te atoms (referred to as Te$_{(1)}$ hereafter), five from each of the Bi atoms and six from the middle layer Te atoms (Te$_{(2)}$).
Whereas Bi-Te$_{(1)}$ bond is fairly ionic in its nature, Bi-Te$_{(2)}$ bond is more covalent and weaker \cite{greenaway1965band}. It is thus expected that when one substitutes Te with a more electronegative Se or S atoms, the more weakly bound Te$_{(2)}$ site is energetically preferred, as shown in Fig.\ref{FIG:Structure}. To get a deeper insight into the electronic structure of the studied monolayers, we use the electron localization function (ELF)\cite{ELF1}, which is known to be a potent tool to identify the localization of electrons and the relative orientation of lone electron pairs. The calculated ELF in Figure~\ref{FIG:Structure}(d) reveals the mushroom shape around the outer layer Te$_{(1)}$ atoms, which evidence the presence of a stereochemically active anionic lone pair with its relative orientation. Our finding is consistent with the previous studies on Bi$_{2}$Te$_{3}$~\cite{Walsh2011,Witting2019,Lee2017} which shows that the anionic lone pair~\cite{Walsh2011,DaSilva2008} with the $p$-state character~\cite{Lee2017} forms on Te$_{(1)}$ atoms, while lone pair Bi-$s$ states are stereochemically inactive~\cite{Walsh2011,Witting2019}.

%\bred{As in outline, start from crystal structure, describe the composition of layers , talk a bout possibility of substitution of the middle layer (WHY not outer layer!!!), and then ELF and lone electron pair as in Figure 1. Leave the electronic and phononic part for their sections.}}
%%%%%%%%%%%%%%%%%%%%%%%%%%%%%%%%%%%%%%%%%%%%%%%%%%%%%%%%%%%%%%%%%%%%%%%%%%%%%%%%%
\vspace{1em}
\noindent{\textbf{Phonon Transport Properties.}} Principles to find materials with intrinsically low lattice thermal conductivity, which was originally proposed by Slack~\cite{slack1973nonmetallic,lindsay2013first,naghavi2018pd2se3,peng2016first}, include, (i) having high average atomic mass, (ii) weak inter-atomic bonding, (iii) complex crystal structure, and (iv) strong anharmonicity, measured by Gr\"{u}neisen parameter.

In many materials including monolayers, a high average mass ($M_{a_v}$) leads to a relatively low Debye temperature ($\Theta_{\rm D}$) and consequently low acoustic phonon velocities \cite{lindsay2013first}. According to this insight, a lower lattice thermal conductivity in Bi$_{2}$Te$_{3}$ is expected compared to that of Sb$_{2}$Te$_{3}$, majorly because of the heavier atomic mass of Bi compared to Sb. 
On the other hand, the bond strength also plays an important role in determining the lattice thermal conductivity, \textit{e.g.}, for monolayers of Mo and W dichalcogenides \cite{huang2015roles,peng2016first}, wherein increasing (decreasing) cation (anion) nucleon number leads to stronger (weaker) bonding and higher (lower) thermal conductivity.
In this regard, the projected Crystal Orbital Overlap Population (pCOOP) and projected Crystal Orbital Hamilton Population (pCOHP) are potent tools for analyzing chemical bonds in solid~\cite{Deringer_2011,Dronskowski_1993,Hoffmann_1987}. The calculated pCOOP and pCOHP are shown in Fig.\,S8 of SI. The integrated values with respect to energy up to $\epsilon_{F}$ for both IpCOOP and IpCOHP can be used to analyze the bond strength~\cite{maintz2016lobster,maintz2013analytic}: the more overlap the orbitals have, the stronger the associated bonds are.  The calculated value for Sb--Te bond in Sb$_{2}$Te$_{2}$Se is about 14\% higher than that of Bi--Te bond in Bi$_{2}$Te$_{2}$Se monolayers. Therefore according to Slack's theory~\cite{slack1973nonmetallic}, the lower thermal conductivity of Bi$_{2}$Te$_{2}$Se monolayers may be partially due to both weaker interatomic bonds and the heavier atomic mass of Bi compared to Sb.

Figure~\ref{FIG:Phonon}(a) shows that Sb$_{2}$Te$_{2}$Se monolayer is dynamically stable with no imaginary modes through the whole BZ. The corresponding atom-decomposed phonon density of states (DOS) shows that in frequencies ranging from 0 to just below 60~cm$^{-1}$---the range which contributes the most to lattice heat transport---the contribution of Sb and Te atoms to DOS are similar. The quantitative description of $\kappa_{\rm L}$ for the studied monolayers is shown in Figure~\ref{FIG:Phonon}(b). The $\kappa_{\rm L}$ of Sb$_{2}$Te$_{2}$Se is 6 Wm$^{-1}$K$^{-1}$ at 300\,K, much lower than the well-studied TE monolayers such as MoS$_{2}$ and WSe$_{2}$, with the $\kappa_{\rm L}$ of 140 and 42 Wm$^{-1}$K$^{-1}$, respectively~\cite{kumar2015thermoelectric}.  
The calculated atom-decomposed lattice thermal conductivity shows that Sb and Te atoms are equally contributing to $\kappa_{\rm L}$ (see Figure~\ref{FIG:Phonon}(c)), consistent with our earlier analysis of the DOS. Moreover,  branch decomposed thermal conductivity (Fig.\,S5, SI) demonstrates that the optical branches contribute poorly to $\kappa_{L}$, hardly reaching 18\% at high temperatures. The largest contributions belong respectively to transverse acoustic (TA), longitudinal acoustic (LA), and out-of-plane acoustic (ZA) branches. The analysis of the mean free path cumulative lattice thermal conductivity (Fig.\,S7, SI) reveals that  the $\kappa_{\rm L}$ can be further suppressed by decreasing the grain size of the polycrystals; for example, at the size of 100\,nm the $\kappa_{\rm L}$ reduces by 50\%. 

Gr\"{u}neisen parameter ($\gamma$) measures the anharmonicity of crystalline systems and based on Slack's theory~\cite{slack1973nonmetallic} it is inversely ($\kappa_{\rm L} \propto \frac{1}{\gamma^{2}})$ related to the lattice thermal conductivity, $\kappa_{\rm L}$. The larger the $\gamma$ the stronger the anharmonicity and thus the lower the $\kappa_{\rm L}$. The Gr\"{u}neisen parameters of Sb$_{2}$Te$_{2}$Se and MoS$_{2}$ are shown in Fig.\,S6(d) and for the transition metal dichalcogenides (TMDs) are reported in Ref.\cite{Peng_2016}. The calculated  Gr\"{u}neisen parameters of acoustic and low-lying optical modes of Sb$_{2}$Te$_{2}$Se ({\it i.e.,}, heat-carrying phonons) evidence not only much higher mode values than MoS$_{2}$ and other TMDs but also stronger $q$-dependence, rendering an intrinsic and large anharmonicity in Sb$_{2}$Te$_{2}$Se. As expected, the $\kappa_{\rm L}$ of Sb$_{2}$Te$_{2}$Se at 300\,K ($\approx$ 6\,Wm$^{-1}$K$^{-1}$) is more than 7 times lower than that of WSe$_{2}$ ($\approx$ 42\,Wm$^{-1}$K$^{-1}$~\cite{Gandi2016}) while the atomic mass of W (183.84 amu) is larger than Sb (121.76 amu)  and Te (127.6 amu). As discussed in the following paragraph, the origin of anharmonicity in Sb$_{2}$Te$_{3}$ is traced back to anionic lone pair of Te atoms.

To find out the fundamental origin of the anharmonicity in Sb$_{2}$Te$_{2}$Se, we calculate the norm of the third-order interatomic force constants (IFCs), which is defined as $\Phi_{mnl}$ = $\frac{\partial^{3}E}{\partial u_{m} \partial u_{n} \partial u_{l}}$ ($E$ and $u$ are the total energy and atom displacement for different atom species $m$, $n$, and $l$). Since (i) a large absolute value of $|\Phi|$ suggests strong anharmonicity and (i) the phonon scattering rates are roughly proportional to $|\Phi_{mnl}|^2$, it can be seen in Figure~\ref{FIG:Phonon}(d) that the strongest anharmonicity is associated with the $\Phi_{TeTeTe}$, namely, the outer layer Te atoms where the lone pair electrons are stereochemically active. Consistent with previous studies~\cite{Dutta2019,Nielsen2013}, the presence of lone pair electrons, localized on Te atoms, could enforce anharmonicity and subsequently lower the lattice thermal conductivity of a crystal.
\begin{figure}[htp!]
	\centering
	\includegraphics[width=1.0\linewidth]{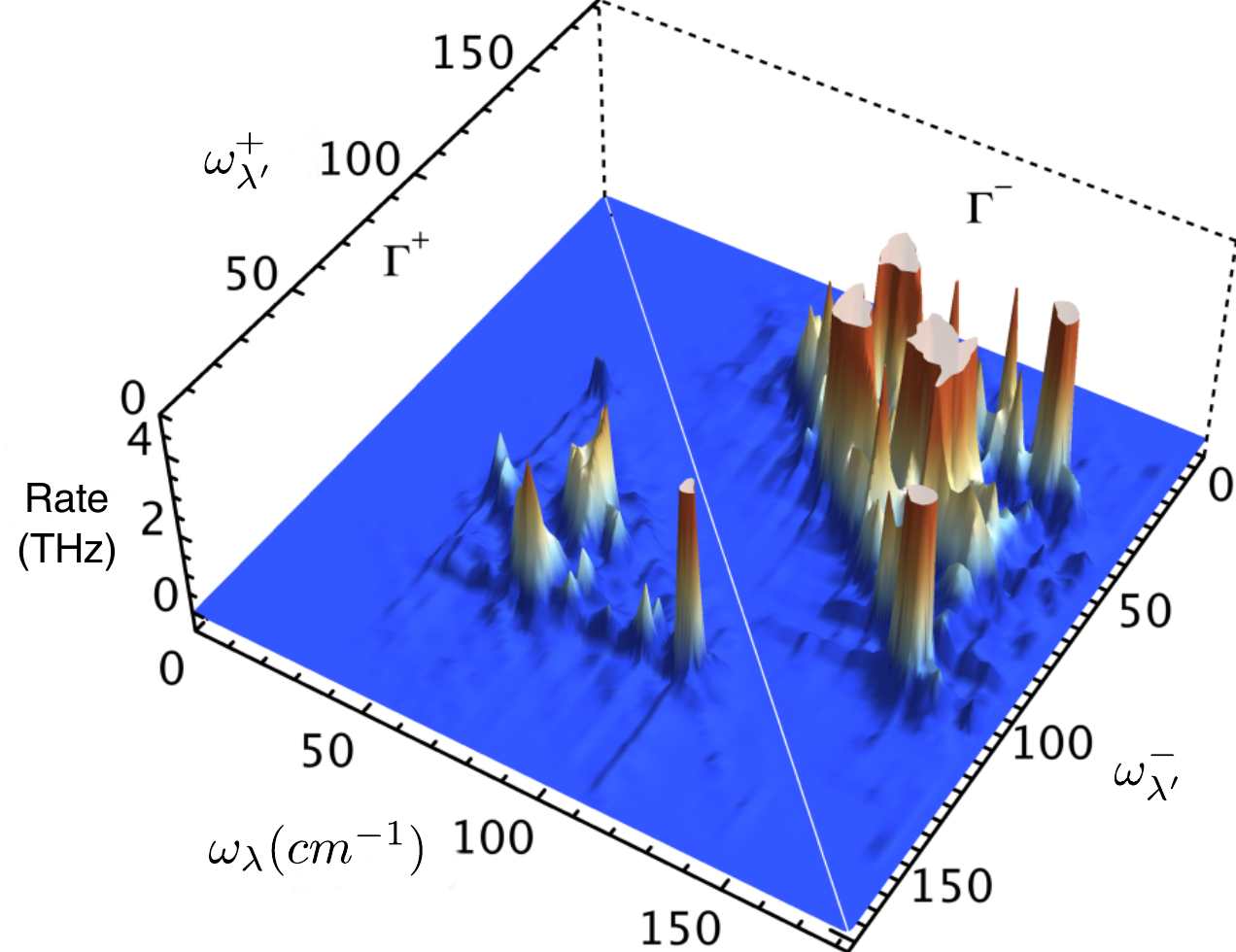}
	\caption{Contour plot of the three-phonon scattering rates associated with the absorption process ($\Gamma^{+}$: $\lambda$ + $\lambda^{'}$ $\rightarrow$ $\lambda^{''}$) and the emission process ($\Gamma^{-}$: $\lambda$ $\rightarrow$  $\lambda^{'}$ + $\lambda^{''}$) in Sb$_{2}$Te$_{2}$Se at 300\,K. The left segment shows the absorption rates, whereas the right panel indicates the rates for emission processes. The corresponding phonon frequencies $\omega$ are given in units of cm$^{-1}$.}
\label{FIG:Emission}
\end{figure}

To further shed light on how phonons are scattered in Sb$_{2}$Te$_{2}$Se, we show in Figure~\ref{FIG:Emission} the frequency-dependent scattering rates arising from the three-phonon interactions, namely, the absorption ($\Gamma_{+}$: $\lambda$ + $\lambda^{'}$ $\rightarrow$ $\lambda^{''}$) and emission ($\Gamma_{-}$: $\lambda$ $\rightarrow$  $\lambda^{'}$ + $\lambda^{''}$) processes. The peaks in the scattering rates plot depict the scattering magnitude of the first phonon mode ($\lambda$) induced by the second phonon mode ($\lambda^{'}$), satisfying both energy and crystal momentum conservation. It can be seen that the low-frequency phonon modes ($<$ 50 cm$^{-1}$) are only weakly scattered, which majorly originates from the absorption process, consistent with their long lifetimes as shown in Fig.S6(c). In contrast, the optical phonon modes are strongly scattered during various emission processes, resulting in small contributions to lattice heat conduction.

% In the absorption process, a low-frequency phonon mode is combined with other low-frequency phonon modes, giving rise to a high-frequency optical mode. In the emission process, a phonon mode is only allowed to decompose into two low-frequency modes, thus restricting the second phonon mode ($\lambda^{'}$).

\begin{figure}[htp!]
	\centering
	\includegraphics[width=1.0\linewidth]{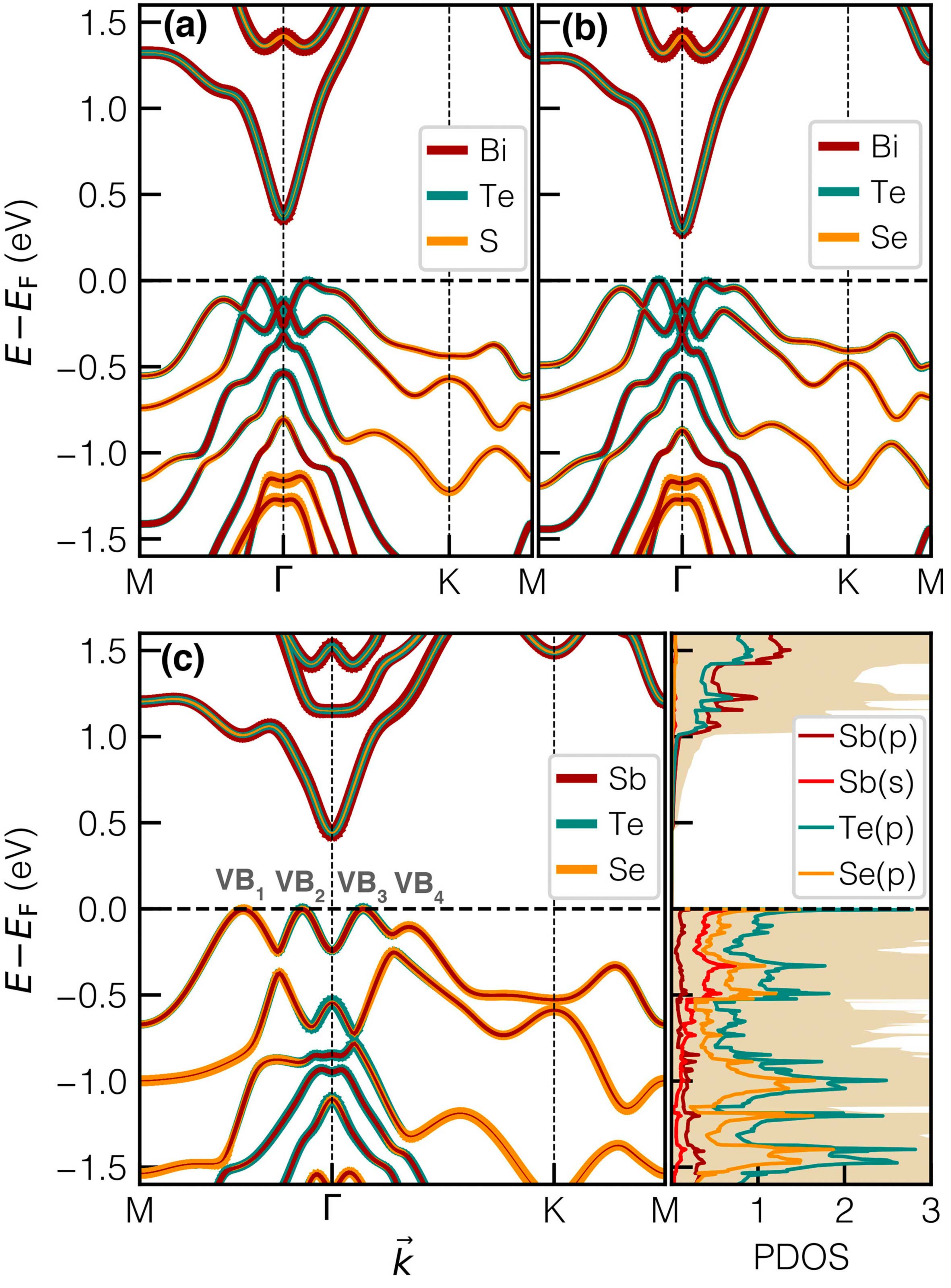}
	\caption{Atom-decomposed electronic band structures of (a) Bi$_{2}$Te$_{2}$S, (b) Bi$_{2}$Te$_{2}$Se, and (c) Sb$_{2}$Te$_{2}$Se monolayers. Considering the similar electronic properties among these three compounds, here we only plot the partial density of states (PDOS) of Sb$_{2}$Te$_{2}$Se and also remove Te(s) and Se(s) projection due to their negligible contributions (see SI for more information). The valence bands are mixed of, in order of magnitude, Te (p), Se/S (p) and Sb/Bi (s) orbitals.}
\label{FIG:Band}
\end{figure}

%%%%%%%%%%%%%%%%%%%%%%%%%%%%%%%%%%%%%%%%%%%%%%%%%%%%%%%%%%%%%%%%%%%%%%%%%%%%%%%%%
\vskip 0.5cm
\noindent{\textbf{Electron Transport Properties.}}
Considering the similarity in the electronic structure of the studied materials, we take Sb$_{2}$Te$_{2}$Se as an example here. As shown in Figure \ref{FIG:Band}, Sb$_{2}$Te$_{2}$Se is a semiconductor with an indirect band-gap of 0.44\,eV (including spin orbit coupling). Figure~\ref{FIG:Band}(c) (see Fig.\,S3, SI, for all family members) show nearly significant contributions of chalcogen and pnictide atoms in both valence and conduction bands, which is consistent with the covalent nature of bonding in this family of semiconductors~\cite{Witting2019,Lee2017,Deringer_2015}. Note that the per atom contributions from Se and Te in the calculated total DOS are similar, however, there are two times more Te atoms than Se in the unit cell, and thus a larger contribution from Te atoms in the valence band is expected.

% In Sb$_{2}$Te$_{2}$Se, Sb loses its 6$p$ electrons to more electronegative atoms, {\it i.e.,} Te and Se, and becomes Sb$^{3+}$. Besides, its 6$s^{2}$ electrons lie well below 6$p$ orbitals of Te and Se atoms, hence, they are stereo-chemically inactive. This tendency for quenching can also be illustrated in terms of high local symmetry, as was shown in the ELF (see Figure~\ref{FIG:Structure}(d). Nevertheless, either being quenched or expressed, it is undeniable that these lone pairs play a major role in the electronic structure of these materials. For instance, the possible explanation for the larger band gap size of Sb$_{2}$Te$_{2}$Se with respect to Bi$_{2}$Te$_{2}$Se (see Figure~\ref{FIG:Band}) is the more energetic behavior of the $s^2$ lone pair in the smaller Sb atoms compared to Bi.

What turns this family into outstanding and the most studied TE materials~\cite{hung2019designing,Sharma_2016,Lee_2018,Zhou_2015,Xu_2018}, beside low lattice thermal conductivity, is their excellent electronic transport properties and relatively high $PF$ compared to other single layer TE candidates.
The feature in the band structure of these monolayers that enhances their $PF$ is the high band degeneracy $N_{\rm v}$, which is usually observed when several valleys with the same energies occur near the Fermi level or when the valley is located at a low symmetry point of the Brillouin zone of a high symmetry lattice (valley degeneracy). For the $R3\bar{m}$ symmetry, valley degeneracy between both $\Gamma\!\!-\!\!M$ and $\Gamma\!\!-\!\!K$ is 6 and thus the $N_{\rm v}$ is just six times the number of energetically degenerate valence band maximum (VBM) at different points in the Brillouin zone. The degeneracy that could occur among VBMs is indeed numerical, also called accidental~\cite{Garrity2016} or band convergence~\cite{tang2015convergence}. Therefore, the energy differences between four potential extrema (see Fig.~\ref{FIG:Band}(c) for VB$_{1}$,VB$_{2}$,VB$_{3}$,VB$_{4}$) with respect to VBM is listed in Table~S1, SI. As discussed in previous studies~\cite{THung2019,hung2019designing}, $PF$ has an exponential dependent on $-\Delta E$, $PF\propto e^{-\Delta E/k_{\rm B}T}$, reaching its maximum value at $\Delta E\approx 0$. As seen in Table~S1 the energy difference ($\Delta E$) between VBM (here VB$_{1}$) and the next highest VB (here VB$_{4}$) of Bi$_{2}$Te$_{3}$ and Sb$_{2}$Te$_{3}$ are 16 and 42\,meV respectively, leaving one VBM at Fermi level, and leading to $N_{\rm v}\!=\!6$. Although the other three valleys are just slightly lower in energy, $N_{\rm v}$ larger than 12 is not reachable by strain engineering. That is because when strain is induced, different valleys in the band structure counteract and raising one results in lowering the others, making further tuning of the valence band insurmountable. For instance, defining $\epsilon = (a-a_{0})/a$ as a measure of change in lattice constant, our calculations for the case of Bi$_{2}$Te$_{3}$ demonstrated that only 4\% strain doubled $N_{\rm v}$ from 6 to 12 and increased the $PF$ by one-tenth (see Fig.\,S1, SI), but this would be an upper limit to $N_{\rm v}$.

%%%%%%%%%%%%%%%%%%%%%%%%%%%%%%%%%%%%%%%%%%%%%%%%%%%%%%%%%%%%%
\vskip 0.25cm

Substitution of Te$_{(2)}$ atoms with S and Se adequately improves band convergence of Bi$_{2}$Te$_{3}$ and Sb$_{2}$Te$_{3}$ monolayers by lowering the $\Delta E$ to $\sim\!\!4$\,meV which can be considered as almost degenerate extrema approaching $N_{\rm v}\!=\!12$. The giant band degeneracy of 18 occurs in Sb$_{2}$Te$_{2}$Se monolayers in which the two band extrema in $\Gamma$--$M$ direction and the one in $\Gamma$--$K$ direction are well-aligned (with numerical differences $<\!3$\,meV). To further understand how atom substitution leads to band alignment, we resort to atom resolved band structure calculations. 

The atom decomposed band structure reveals that beside Sb/Bi atoms, the outer layer chalcogen atoms (mainly their $p$-state as shown in DOS of Figure~\ref{FIG:Band}(c)) are significantly contributing to VB$_{2}$ and VB$_{3}$: that is to say, these extrema mainly inherit the character of the Te$_{(1)}$ atoms (bold green line). On the contrary, in the VB$_{1}$ and VB$_{4}$, the middle layer chalcogen atoms ({\it i.e.,} Te$_{(2)}$ have the dominant contribution (bold orange line). Therefore, one can independently tune the height of VB$_{1,4}$ with respect to VB$_{2,3}$ by changing the middle layer atoms. As seen in Fig.S2, SI, substitution of the middle layer Te$_{(2)}$ atoms with the more electronegative Se ones in Sb$_{2}$Te$_{3}$ lowers the height of VB$_{1,4}$ with respect to the VB$_{2,3}$ , which leads to the alignment of VB$_{1,2,3}$, and to $N_{\rm v}=18$ in Sb$_{2}$Te$_{2}$Se as highlighted in Figure~\ref{FIG:Band}(c). The same observation holds  for S substituted monolayers, however, due to higher electronegativity of S compared to Se the VB$_{1}$ and VB$_{4}$ extrema decrease a lot that leave only the two central extrema aligned, leading to $N_{\rm v}=12$ in Sb$_{2}$Te$_{2}$S.

Using a Constant Relaxation Time Approximation (CRTA), we calculated electronic transport coefficients for Bi$_{2}$Te$_{2}$S, Bi$_{2}$Te$_{2}$Se and Sb$_{2}$Te$_{2}$Se monolayers at varying temperatures by solving the Boltzmann transport equation. 
To estimate electronic transport properties of materials the choice of electron relaxation time $\tau$ is critical. From comparison with experimental data at 300\,K, N.\,F.\,Hinsche {\it{et\,al.}}~\cite{Hinsche2011} determined the in-plane $\tau$=$1.2\!\times\!10^{-14}$\,s and $1.1\!\times\!10^{-14}$\,s for the Sb$_{2}$Te$_{3}$ and Bi$_{2}$Te$_{3}$, respectively. Also, fitted values such as $2.2\!\times\!10^{-14}$\,s~\cite{Scheidemantel2003,Zhou_2015} and $\approx 0.65\!\times\!10^{-14}$~\cite{Lee_2018} were suggested for Bi$_{2}$Te$_{3}$ at 300\,K. Likewise, for Bi$_{2}$Te$_{2}$S the in-plane relaxation time of $3\!\times\!10^{-14}$\,s was used~\cite{Joo2019}. N. T. Hung {\it{et\,al.,}}~~\cite{hung2019designing} calculated the relaxation time $\tau$ in Bi$_{2}$\{S, Se, Te\}$_{3}$ monolayers and found it to be $0.2-0.5\!\times\!10^{-14}$\,s, much lower than their 3D counterparts. They claimed that shorter $\tau$ for 2D materials is due to quantum confinement effect~\cite{Hicks1993} which increases the density of electronic states in 2D systems. Here we use $\tau\!=\!0.5\!\times\!10^{-14}$\,s and $1\!\times\!10^{-14}$\,s, as suggested by Ref.~\cite{Hinsche2011} and Ref.~\cite{hung2019designing} to estimate the $zT$ of the studied materials. However, because of possible uncertainties in the estimation of $\tau$ and to make the comparison to previous work easier,  we also report the $zT$ and $PF$ for different $\tau$ (color bar) within a range from 0.2 to 4.4$\times\!10^{-14}$\,s in Fig.\,S9 of SI, covering all the previously reported $\tau$.

\begin{figure}[htp!]
	\centering
	\includegraphics[width=1.0\linewidth]{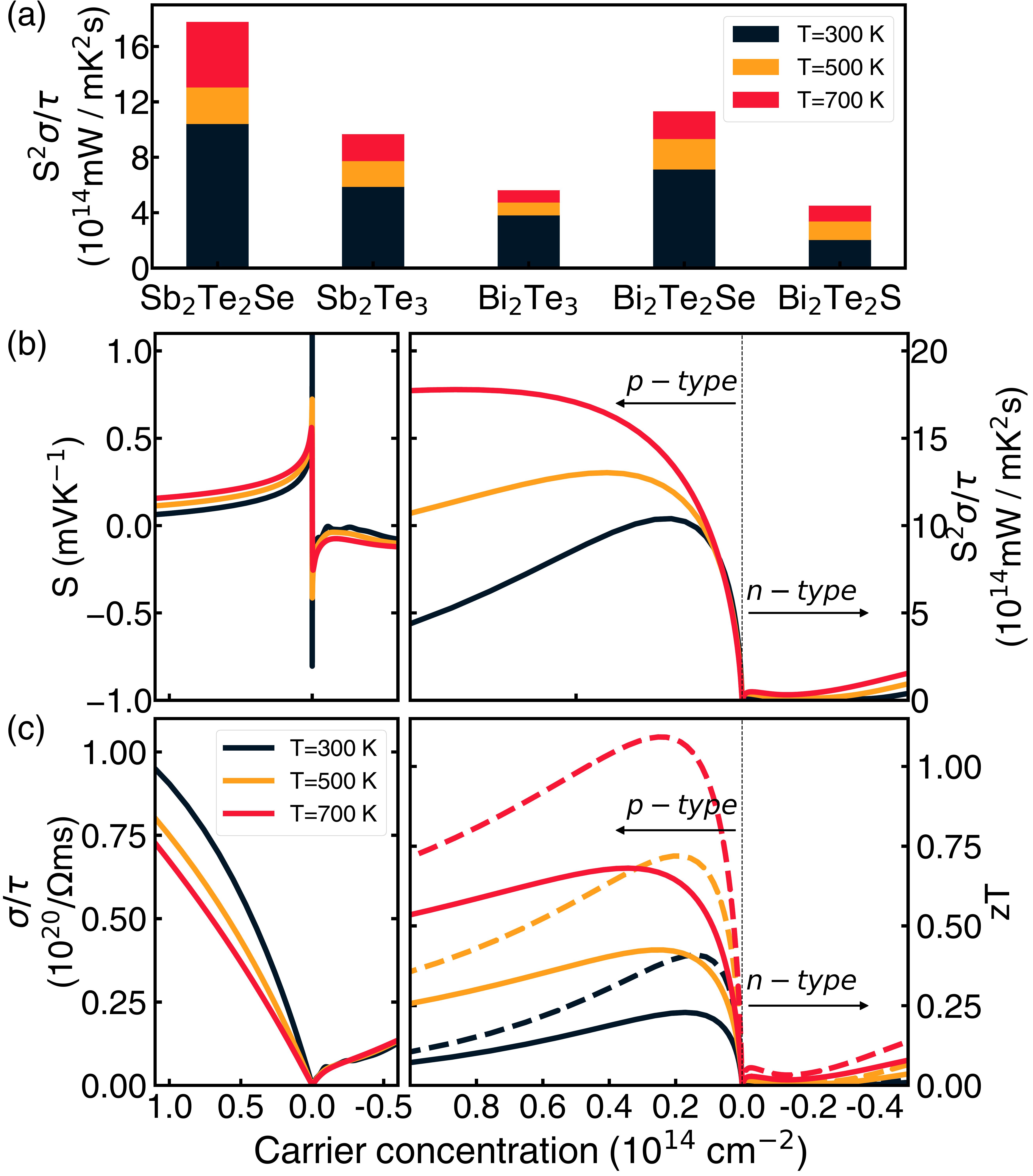}
	\caption{(a) Optimal PF of different monolayer compounds are presented at 300, 500 and 700\,K for the purpose of comparison. The figure indicates a PF increase in newly proposed materials, Sb$_{2}$Te$_{2}$Se and Bi$_{2}$Te$_{2}$Se, with respect to their parents. (b) Transport coefficients (PF, S), and (c) $\sigma$ and $zT$ of Sb$_{2}$Te$_{2}$Se as a function of carrier concentrations as functions of temperatures and relaxation times. Solid and dashed lines represent $zT$ values associated with $\tau$=0.5 and 1.0\,$\times\!10^{-14}$~s, respectively.}
\label{FIG:ZT}
\end{figure}

Assuming  $\tau\!\!=\!\!1\times\!10^{-14}$\,s, the calculated $PF$ of Sb$_{2}$Te$_{2}$Se at three different temperatures are  compared to other studied compounds. Note that the increase in $PF$ at higher temperatures is due to a constant relaxation time adopted in all three temperatures, while in reality the relaxation time decreases with increasing temperature (roughly by $\tau\!\propto\!T^{-1.5}$ for Bi$_{2}$Te$_{3}$~\cite{Zhou_2015}). As seen in Figure~\ref{FIG:ZT}(a), using $\tau\!\!=\!\!1\times\!10^{-14}$\,s at $T\!\!=300$\,K our calculated $PF$ for Bi$_{2}$Te$_{3}$ is about 4~$\rm mWm^{-1}K^{-2}$ that agrees well with the $\sim 4.4$ $\rm mWm^{-1}K^{-2}$ reported in the previous study~\cite{Zhou_2015}. This $PF$ increases to 7~$\rm mWm^{-1}K^{-2}$ by substitution of Te with Se atom as  in Bi$_{2}$Te$_{2}$Se. Likewise, the $PF$ of Sb$_{2}$Te$_{3}$ increases by a factor of two upon Se substitution, achieving a giant $PF\!\!=\!10$~$\rm mWm^{-1}K^{-2}$ in Sb$_{2}$Te$_{2}$Se at 300\,K as seen in Figure~\ref{FIG:ZT}(b). Using the same $\tau$ and temperature, the maximum $PF$ for MoS$_{2}$, MoSe$_{2}$, TiS$_{3}$, Pd$_{2}$Se$_{3}$ are respectively 1.8~$\rm mWm^{-1}K^{-2}$~\cite{babaei2014large} (n-type), 0.8~$\rm mWm^{-1}K^{-2}$~\cite{kumar2015thermoelectric} (n-type) and 1.8~$\rm mWm^{-1}K^{-2}$~\cite{Zhang2017} (n-type) and 1.6~$\rm mWm^{-1}K^{-2}$~\cite{naghavi2018pd2se3} (p-type). Due to the much larger $PF$ and much smaller $\kappa_{\rm L}$ compared to these materials (\textit{e.g.}, $\kappa_{\rm L}$ of Sb$_{2}$Te$_{2}$Se at 300\,K is 6~Wm$^{-1}$K$^{-1}$ and MoS$_{2}$ 131~Wm$^{-1}$K$^{-1}$~\cite{Gandi2016}) we expect a high figure of merit in Sb$_{2}$Te$_{2}$Se, thus enabling high thermoelectric efficiency.
%%%%%%%%%%%%%%%%%%%%%%%%%%%%%%%%%%%%%%%%%%%%%%%%%%%%%%
\vskip 0.1cm
%%%%%%%%%%%%%%%%%%%%%%%%%%%%%%%%%%%%%%%%%%%%%%%%%%%%%%%%%%%%%%%%%%%%%%%%%%%%%%%%%
\vspace{0em} 
\noindent{\textbf{Thermoelectric Figure of Merit.}}
Figure~\ref{FIG:ZT}(c) shows the calculated $zT$ for the Sb$_{2}$Te$_{2}$Se at three different relaxation times and temperatures. Even assuming $\tau$ as small as $0.5\!\times\!10^{-14}$\,s we could reach the $zT$ of 0.7 and 0.22 at 700 and 300\,K, respectively.  Considering the same relaxation time, the calculated $zT$ values are much larger than other proposed monolayers such as WSe$_{2}$~\cite{kumar2015thermoelectric} and TiS$_{3}$~\cite{Zhang2017} and SnSe~\cite{Wang2015}. For example, using the same $\tau$ the calculated $zT\!\!=0.7$ at 700\,K for Sb$_{2}$Te$_{2}$Se is 30\% larger than SnSe monolayers~\cite{Wang2015}, which has the highest record $zT$ in the bulk phase~\cite{Zhao2014}.

It is worth nothing that the estimated $zT$ in this work could be considered as a lower limit, because the used electron relaxation time is quite small and the calculated $\kappa_{\rm L}$ tends to be larger than the experimental one. For example, the calculated $\kappa_{\rm L}$ for MoS$_{2}$ monolayers is 131~Wm$^{-1}$K$^{-1}$~\cite{Gandi2016} while experimental value of 34.5~Wm$^{-1}$K$^{-1}$ was measured~\cite{Yan2014}, which could be due to point defects, dislocations, and finite size effect. Therefore, the real $zT$ could be larger than our estimation. 
%%%%%%%%%%%%%%%%%%%%%%%%%%%%%%%%%%%%%%%%%%%%%%%%%%%%%%%%%%%%%%%%%%%%%%%%%%%%%%%%%
\vskip 1cm
\noindent{\textbf{\large{Conclusions}}} \\[0.5em]
We investigated electronic structure, phonon and electron transport properties of $M_{2}$Te$_{2}X$ (M=Bi, Sb; X=S, Se, Te) monolayers by the means of first-principles calculations and Boltzmann transport theory.  Our detailed analysis reveal that four reachable valence band extrema exist in this family of materials and that two of them merely depends on the electronegativity of the middle layer chalcogen atoms, which makes the independent modulation of the vallyes' position doable. As such we substitute the middle layer Te with isoelectronic equivalents S, and Se atoms and found that the band alignment significantly improves transport properties of both Bi$_{2}$Te$_{3}$ and Sb$_{2}$Te$_{3}$. The superior valence band alignment occurs in Sb$_{2}$Te$_{2}$Se monolayers where the trhee out of four possible valleys are well-aligned resulting in the giant band degeneracy of 18 that the hold a record among all thermoelectric materials. Our results demonstrate that $zT$ in this family of monolayers, is comparable to, and can be better than many well-known thermoelectric compounds. It should also be noted that the reported values for $zT$ in these materials are only lower limits, since nanostructuring can significantly deteriorate lattice thermal conductivity by almost 50 percent. Our finding could advance the thermoelectric materials design, where the convergence of electronic band is achieved by tuning the composition.

%%%%% % %%%%%%%%%%%%%%%%%%%%%%%%%%%%%%%%%%%%%%%%%%%%%%%%%%%%%%%%%%%%%%%%%%%%%%%%%%%%%%%%%
%%%%% \begin{acknowledgments} 
%%%%% \end{acknowledgments}
%%%%% % %%%%%%%%%%%%%%%%%%%%%%%%%%%%%%%%%%%%%%%%%%%%%%%%%%%%%%%%%%%%%%%%%%%%%%%%%%%%%%%%%
\clearpage
%\bibliography{references_final}

%merlin.mbs apsrev4-1.bst 2010-07-25 4.21a (PWD, AO, DPC) hacked
%Control: key (0)
%Control: author (8) initials jnrlst
%Control: editor formatted (1) identically to author
%Control: production of article title (-1) disabled
%Control: page (0) single
%Control: year (1) truncated
%Control: production of eprint (0) enabled
\begin{thebibliography}{75}%
\makeatletter
\providecommand \@ifxundefined [1]{%
 \@ifx{#1\undefined}
}%
\providecommand \@ifnum [1]{%
 \ifnum #1\expandafter \@firstoftwo
 \else \expandafter \@secondoftwo
 \fi
}%
\providecommand \@ifx [1]{%
 \ifx #1\expandafter \@firstoftwo
 \else \expandafter \@secondoftwo
 \fi
}%
\providecommand \natexlab [1]{#1}%
\providecommand \enquote  [1]{``#1''}%
\providecommand \bibnamefont  [1]{#1}%
\providecommand \bibfnamefont [1]{#1}%
\providecommand \citenamefont [1]{#1}%
\providecommand \href@noop [0]{\@secondoftwo}%
\providecommand \href [0]{\begingroup \@sanitize@url \@href}%
\providecommand \@href[1]{\@@startlink{#1}\@@href}%
\providecommand \@@href[1]{\endgroup#1\@@endlink}%
\providecommand \@sanitize@url [0]{\catcode `\\12\catcode `\$12\catcode
  `\&12\catcode `\#12\catcode `\^12\catcode `\_12\catcode `\%12\relax}%
\providecommand \@@startlink[1]{}%
\providecommand \@@endlink[0]{}%
\providecommand \url  [0]{\begingroup\@sanitize@url \@url }%
\providecommand \@url [1]{\endgroup\@href {#1}{\urlprefix }}%
\providecommand \urlprefix  [0]{URL }%
\providecommand \Eprint [0]{\href }%
\providecommand \doibase [0]{http://dx.doi.org/}%
\providecommand \selectlanguage [0]{\@gobble}%
\providecommand \bibinfo  [0]{\@secondoftwo}%
\providecommand \bibfield  [0]{\@secondoftwo}%
\providecommand \translation [1]{[#1]}%
\providecommand \BibitemOpen [0]{}%
\providecommand \bibitemStop [0]{}%
\providecommand \bibitemNoStop [0]{.\EOS\space}%
\providecommand \EOS [0]{\spacefactor3000\relax}%
\providecommand \BibitemShut  [1]{\csname bibitem#1\endcsname}%
\let\auto@bib@innerbib\@empty
%</preamble>
\bibitem [{\citenamefont {Goldsmid}(2010)}]{goldsmid2010introduction}%
  \BibitemOpen
  \bibfield  {author} {\bibinfo {author} {\bibfnamefont {H.~J.}\ \bibnamefont
  {Goldsmid}},\ }\href@noop {} {\emph {\bibinfo {title} {Introduction to
  thermoelectricity}}},\ Vol.\ \bibinfo {volume} {121}\ (\bibinfo  {publisher}
  {Springer},\ \bibinfo {year} {2010})\BibitemShut {NoStop}%
\bibitem [{\citenamefont {Tang}\ \emph {et~al.}(2010)\citenamefont {Tang},
  \citenamefont {Wang}, \citenamefont {Lee}, \citenamefont {Fardy},
  \citenamefont {Huo}, \citenamefont {Russell},\ and\ \citenamefont
  {Yang}}]{tang2010holey}%
  \BibitemOpen
  \bibfield  {author} {\bibinfo {author} {\bibfnamefont {J.}~\bibnamefont
  {Tang}}, \bibinfo {author} {\bibfnamefont {H.-T.}\ \bibnamefont {Wang}},
  \bibinfo {author} {\bibfnamefont {D.~H.}\ \bibnamefont {Lee}}, \bibinfo
  {author} {\bibfnamefont {M.}~\bibnamefont {Fardy}}, \bibinfo {author}
  {\bibfnamefont {Z.}~\bibnamefont {Huo}}, \bibinfo {author} {\bibfnamefont
  {T.~P.}\ \bibnamefont {Russell}}, \ and\ \bibinfo {author} {\bibfnamefont
  {P.}~\bibnamefont {Yang}},\ }\href@noop {} {\bibfield  {journal} {\bibinfo
  {journal} {Nano letters}\ }\textbf {\bibinfo {volume} {10}},\ \bibinfo
  {pages} {4279} (\bibinfo {year} {2010})}\BibitemShut {NoStop}%
\bibitem [{\citenamefont {Majumdar}(2004)}]{majumdar2004thermoelectricity}%
  \BibitemOpen
  \bibfield  {author} {\bibinfo {author} {\bibfnamefont {A.}~\bibnamefont
  {Majumdar}},\ }\href@noop {} {\bibfield  {journal} {\bibinfo  {journal}
  {Science}\ }\textbf {\bibinfo {volume} {303}},\ \bibinfo {pages} {777}
  (\bibinfo {year} {2004})}\BibitemShut {NoStop}%
\bibitem [{\citenamefont {Poudel}\ \emph {et~al.}(2008)\citenamefont {Poudel},
  \citenamefont {Hao}, \citenamefont {Ma}, \citenamefont {Lan}, \citenamefont
  {Minnich}, \citenamefont {Yu}, \citenamefont {Yan}, \citenamefont {Wang},
  \citenamefont {Muto}, \citenamefont {Vashaee} \emph
  {et~al.}}]{poudel2008high}%
  \BibitemOpen
  \bibfield  {author} {\bibinfo {author} {\bibfnamefont {B.}~\bibnamefont
  {Poudel}}, \bibinfo {author} {\bibfnamefont {Q.}~\bibnamefont {Hao}},
  \bibinfo {author} {\bibfnamefont {Y.}~\bibnamefont {Ma}}, \bibinfo {author}
  {\bibfnamefont {Y.}~\bibnamefont {Lan}}, \bibinfo {author} {\bibfnamefont
  {A.}~\bibnamefont {Minnich}}, \bibinfo {author} {\bibfnamefont
  {B.}~\bibnamefont {Yu}}, \bibinfo {author} {\bibfnamefont {X.}~\bibnamefont
  {Yan}}, \bibinfo {author} {\bibfnamefont {D.}~\bibnamefont {Wang}}, \bibinfo
  {author} {\bibfnamefont {A.}~\bibnamefont {Muto}}, \bibinfo {author}
  {\bibfnamefont {D.}~\bibnamefont {Vashaee}},  \emph {et~al.},\ }\href@noop {}
  {\bibfield  {journal} {\bibinfo  {journal} {Science}\ }\textbf {\bibinfo
  {volume} {320}},\ \bibinfo {pages} {634} (\bibinfo {year}
  {2008})}\BibitemShut {NoStop}%
\bibitem [{\citenamefont {Dresselhaus}\ \emph {et~al.}(1999)\citenamefont
  {Dresselhaus}, \citenamefont {Dresselhaus}, \citenamefont {Sun},
  \citenamefont {Zhang}, \citenamefont {Cronin},\ and\ \citenamefont
  {Koga}}]{dresselhaus1999low}%
  \BibitemOpen
  \bibfield  {author} {\bibinfo {author} {\bibfnamefont {M.}~\bibnamefont
  {Dresselhaus}}, \bibinfo {author} {\bibfnamefont {G.}~\bibnamefont
  {Dresselhaus}}, \bibinfo {author} {\bibfnamefont {X.}~\bibnamefont {Sun}},
  \bibinfo {author} {\bibfnamefont {Z.}~\bibnamefont {Zhang}}, \bibinfo
  {author} {\bibfnamefont {S.}~\bibnamefont {Cronin}}, \ and\ \bibinfo {author}
  {\bibfnamefont {T.}~\bibnamefont {Koga}},\ }\href@noop {} {\bibfield
  {journal} {\bibinfo  {journal} {Physics of the Solid State}\ }\textbf
  {\bibinfo {volume} {41}},\ \bibinfo {pages} {679} (\bibinfo {year}
  {1999})}\BibitemShut {NoStop}%
\bibitem [{\citenamefont {Liu}\ \emph {et~al.}(2012)\citenamefont {Liu},
  \citenamefont {Tan}, \citenamefont {Yin}, \citenamefont {Liu}, \citenamefont
  {Tang}, \citenamefont {Shi}, \citenamefont {Zhang},\ and\ \citenamefont
  {Uher}}]{liu2012convergence}%
  \BibitemOpen
  \bibfield  {author} {\bibinfo {author} {\bibfnamefont {W.}~\bibnamefont
  {Liu}}, \bibinfo {author} {\bibfnamefont {X.}~\bibnamefont {Tan}}, \bibinfo
  {author} {\bibfnamefont {K.}~\bibnamefont {Yin}}, \bibinfo {author}
  {\bibfnamefont {H.}~\bibnamefont {Liu}}, \bibinfo {author} {\bibfnamefont
  {X.}~\bibnamefont {Tang}}, \bibinfo {author} {\bibfnamefont {J.}~\bibnamefont
  {Shi}}, \bibinfo {author} {\bibfnamefont {Q.}~\bibnamefont {Zhang}}, \ and\
  \bibinfo {author} {\bibfnamefont {C.}~\bibnamefont {Uher}},\ }\href@noop {}
  {\bibfield  {journal} {\bibinfo  {journal} {Physical Review Letters}\
  }\textbf {\bibinfo {volume} {108}},\ \bibinfo {pages} {166601} (\bibinfo
  {year} {2012})}\BibitemShut {NoStop}%
\bibitem [{\citenamefont {Tang}\ \emph {et~al.}(2015)\citenamefont {Tang},
  \citenamefont {Gibbs}, \citenamefont {Agapito}, \citenamefont {Li},
  \citenamefont {Kim}, \citenamefont {Nardelli}, \citenamefont {Curtarolo},\
  and\ \citenamefont {Snyder}}]{tang2015convergence}%
  \BibitemOpen
  \bibfield  {author} {\bibinfo {author} {\bibfnamefont {Y.}~\bibnamefont
  {Tang}}, \bibinfo {author} {\bibfnamefont {Z.~M.}\ \bibnamefont {Gibbs}},
  \bibinfo {author} {\bibfnamefont {L.~A.}\ \bibnamefont {Agapito}}, \bibinfo
  {author} {\bibfnamefont {G.}~\bibnamefont {Li}}, \bibinfo {author}
  {\bibfnamefont {H.-S.}\ \bibnamefont {Kim}}, \bibinfo {author} {\bibfnamefont
  {M.~B.}\ \bibnamefont {Nardelli}}, \bibinfo {author} {\bibfnamefont
  {S.}~\bibnamefont {Curtarolo}}, \ and\ \bibinfo {author} {\bibfnamefont
  {G.~J.}\ \bibnamefont {Snyder}},\ }\href@noop {} {\bibfield  {journal}
  {\bibinfo  {journal} {Nature materials}\ }\textbf {\bibinfo {volume} {14}},\
  \bibinfo {pages} {1223} (\bibinfo {year} {2015})}\BibitemShut {NoStop}%
\bibitem [{\citenamefont {Zhang}\ \emph {et~al.}(2012)\citenamefont {Zhang},
  \citenamefont {Cao}, \citenamefont {Liu}, \citenamefont {Lukas},
  \citenamefont {Yu}, \citenamefont {Chen}, \citenamefont {Opeil},
  \citenamefont {Broido}, \citenamefont {Chen},\ and\ \citenamefont
  {Ren}}]{zhang2012heavy}%
  \BibitemOpen
  \bibfield  {author} {\bibinfo {author} {\bibfnamefont {Q.}~\bibnamefont
  {Zhang}}, \bibinfo {author} {\bibfnamefont {F.}~\bibnamefont {Cao}}, \bibinfo
  {author} {\bibfnamefont {W.}~\bibnamefont {Liu}}, \bibinfo {author}
  {\bibfnamefont {K.}~\bibnamefont {Lukas}}, \bibinfo {author} {\bibfnamefont
  {B.}~\bibnamefont {Yu}}, \bibinfo {author} {\bibfnamefont {S.}~\bibnamefont
  {Chen}}, \bibinfo {author} {\bibfnamefont {C.}~\bibnamefont {Opeil}},
  \bibinfo {author} {\bibfnamefont {D.}~\bibnamefont {Broido}}, \bibinfo
  {author} {\bibfnamefont {G.}~\bibnamefont {Chen}}, \ and\ \bibinfo {author}
  {\bibfnamefont {Z.}~\bibnamefont {Ren}},\ }\href@noop {} {\bibfield
  {journal} {\bibinfo  {journal} {Journal of the American chemical society}\
  }\textbf {\bibinfo {volume} {134}},\ \bibinfo {pages} {10031} (\bibinfo
  {year} {2012})}\BibitemShut {NoStop}%
\bibitem [{\citenamefont {Witkoske}\ \emph {et~al.}(2017)\citenamefont
  {Witkoske}, \citenamefont {Wang}, \citenamefont {Lundstrom}, \citenamefont
  {Askarpour},\ and\ \citenamefont {Maassen}}]{witkoske2017thermoelectric}%
  \BibitemOpen
  \bibfield  {author} {\bibinfo {author} {\bibfnamefont {E.}~\bibnamefont
  {Witkoske}}, \bibinfo {author} {\bibfnamefont {X.}~\bibnamefont {Wang}},
  \bibinfo {author} {\bibfnamefont {M.}~\bibnamefont {Lundstrom}}, \bibinfo
  {author} {\bibfnamefont {V.}~\bibnamefont {Askarpour}}, \ and\ \bibinfo
  {author} {\bibfnamefont {J.}~\bibnamefont {Maassen}},\ }\href@noop {}
  {\bibfield  {journal} {\bibinfo  {journal} {Journal of Applied Physics}\
  }\textbf {\bibinfo {volume} {122}},\ \bibinfo {pages} {175102} (\bibinfo
  {year} {2017})}\BibitemShut {NoStop}%
\bibitem [{\citenamefont {Pei}\ \emph {et~al.}(2011)\citenamefont {Pei},
  \citenamefont {Shi}, \citenamefont {LaLonde}, \citenamefont {Wang},
  \citenamefont {Chen},\ and\ \citenamefont {Snyder}}]{pei2011convergence}%
  \BibitemOpen
  \bibfield  {author} {\bibinfo {author} {\bibfnamefont {Y.}~\bibnamefont
  {Pei}}, \bibinfo {author} {\bibfnamefont {X.}~\bibnamefont {Shi}}, \bibinfo
  {author} {\bibfnamefont {A.}~\bibnamefont {LaLonde}}, \bibinfo {author}
  {\bibfnamefont {H.}~\bibnamefont {Wang}}, \bibinfo {author} {\bibfnamefont
  {L.}~\bibnamefont {Chen}}, \ and\ \bibinfo {author} {\bibfnamefont {G.~J.}\
  \bibnamefont {Snyder}},\ }\href@noop {} {\bibfield  {journal} {\bibinfo
  {journal} {Nature}\ }\textbf {\bibinfo {volume} {473}},\ \bibinfo {pages}
  {66} (\bibinfo {year} {2011})}\BibitemShut {NoStop}%
\bibitem [{\citenamefont {Pei}\ \emph {et~al.}(2012)\citenamefont {Pei},
  \citenamefont {Wang},\ and\ \citenamefont {Snyder}}]{pei2012band}%
  \BibitemOpen
  \bibfield  {author} {\bibinfo {author} {\bibfnamefont {Y.}~\bibnamefont
  {Pei}}, \bibinfo {author} {\bibfnamefont {H.}~\bibnamefont {Wang}}, \ and\
  \bibinfo {author} {\bibfnamefont {G.~J.}\ \bibnamefont {Snyder}},\
  }\href@noop {} {\bibfield  {journal} {\bibinfo  {journal} {Advanced
  materials}\ }\textbf {\bibinfo {volume} {24}},\ \bibinfo {pages} {6125}
  (\bibinfo {year} {2012})}\BibitemShut {NoStop}%
\bibitem [{\citenamefont {He}\ \emph {et~al.}(2017)\citenamefont {He},
  \citenamefont {Hao}, \citenamefont {Xia}, \citenamefont {Naghavi},
  \citenamefont {Ozoliņš},\ and\ \citenamefont
  {Wolverton}}]{he2017bi2pdo4}%
  \BibitemOpen
  \bibfield  {author} {\bibinfo {author} {\bibfnamefont {J.}~\bibnamefont
  {He}}, \bibinfo {author} {\bibfnamefont {S.}~\bibnamefont {Hao}}, \bibinfo
  {author} {\bibfnamefont {Y.}~\bibnamefont {Xia}}, \bibinfo {author}
  {\bibfnamefont {S.~S.}\ \bibnamefont {Naghavi}}, \bibinfo {author}
  {\bibfnamefont {V.}~\bibnamefont {Ozoliņš}}, \ and\ \bibinfo {author}
  {\bibfnamefont {C.}~\bibnamefont {Wolverton}},\ }\href@noop {} {\bibfield
  {journal} {\bibinfo  {journal} {Chemistry of Materials}\ }\textbf {\bibinfo
  {volume} {29}},\ \bibinfo {pages} {2529} (\bibinfo {year}
  {2017})}\BibitemShut {NoStop}%
\bibitem [{\citenamefont {He}\ \emph {et~al.}(2019)\citenamefont {He},
  \citenamefont {Xia}, \citenamefont {Naghavi}, \citenamefont
  {Ozoli{\c{n}}{\v{s}}},\ and\ \citenamefont {Wolverton}}]{he2019designing}%
  \BibitemOpen
  \bibfield  {author} {\bibinfo {author} {\bibfnamefont {J.}~\bibnamefont
  {He}}, \bibinfo {author} {\bibfnamefont {Y.}~\bibnamefont {Xia}}, \bibinfo
  {author} {\bibfnamefont {S.~S.}\ \bibnamefont {Naghavi}}, \bibinfo {author}
  {\bibfnamefont {V.}~\bibnamefont {Ozoli{\c{n}}{\v{s}}}}, \ and\ \bibinfo
  {author} {\bibfnamefont {C.}~\bibnamefont {Wolverton}},\ }\href@noop {}
  {\bibfield  {journal} {\bibinfo  {journal} {Nature communications}\ }\textbf
  {\bibinfo {volume} {10}},\ \bibinfo {pages} {719} (\bibinfo {year}
  {2019})}\BibitemShut {NoStop}%
\bibitem [{\citenamefont {Goldsmid}(2013)}]{goldsmid2013thermoelectric}%
  \BibitemOpen
  \bibfield  {author} {\bibinfo {author} {\bibfnamefont {H.}~\bibnamefont
  {Goldsmid}},\ }\href@noop {} {\emph {\bibinfo {title} {Thermoelectric
  refrigeration}}}\ (\bibinfo  {publisher} {Springer},\ \bibinfo {year}
  {2013})\BibitemShut {NoStop}%
\bibitem [{\citenamefont {Wright}(1958)}]{wright1958thermoelectric}%
  \BibitemOpen
  \bibfield  {author} {\bibinfo {author} {\bibfnamefont {D.}~\bibnamefont
  {Wright}},\ }\href@noop {} {\bibfield  {journal} {\bibinfo  {journal}
  {Nature}\ }\textbf {\bibinfo {volume} {181}},\ \bibinfo {pages} {834}
  (\bibinfo {year} {1958})}\BibitemShut {NoStop}%
\bibitem [{\citenamefont {Kim}\ \emph {et~al.}(2017)\citenamefont {Kim},
  \citenamefont {Heinz}, \citenamefont {Gibbs}, \citenamefont {Tang},
  \citenamefont {Kang},\ and\ \citenamefont {Snyder}}]{kim2017high}%
  \BibitemOpen
  \bibfield  {author} {\bibinfo {author} {\bibfnamefont {H.-S.}\ \bibnamefont
  {Kim}}, \bibinfo {author} {\bibfnamefont {N.~A.}\ \bibnamefont {Heinz}},
  \bibinfo {author} {\bibfnamefont {Z.~M.}\ \bibnamefont {Gibbs}}, \bibinfo
  {author} {\bibfnamefont {Y.}~\bibnamefont {Tang}}, \bibinfo {author}
  {\bibfnamefont {S.~D.}\ \bibnamefont {Kang}}, \ and\ \bibinfo {author}
  {\bibfnamefont {G.~J.}\ \bibnamefont {Snyder}},\ }\href@noop {} {\bibfield
  {journal} {\bibinfo  {journal} {Materials Today}\ }\textbf {\bibinfo {volume}
  {20}},\ \bibinfo {pages} {452} (\bibinfo {year} {2017})}\BibitemShut
  {NoStop}%
\bibitem [{\citenamefont {Yan}\ \emph {et~al.}(2010)\citenamefont {Yan},
  \citenamefont {Poudel}, \citenamefont {Ma}, \citenamefont {Liu},
  \citenamefont {Joshi}, \citenamefont {Wang}, \citenamefont {Lan},
  \citenamefont {Wang}, \citenamefont {Chen},\ and\ \citenamefont
  {Ren}}]{yan2010experimental}%
  \BibitemOpen
  \bibfield  {author} {\bibinfo {author} {\bibfnamefont {X.}~\bibnamefont
  {Yan}}, \bibinfo {author} {\bibfnamefont {B.}~\bibnamefont {Poudel}},
  \bibinfo {author} {\bibfnamefont {Y.}~\bibnamefont {Ma}}, \bibinfo {author}
  {\bibfnamefont {W.}~\bibnamefont {Liu}}, \bibinfo {author} {\bibfnamefont
  {G.}~\bibnamefont {Joshi}}, \bibinfo {author} {\bibfnamefont
  {H.}~\bibnamefont {Wang}}, \bibinfo {author} {\bibfnamefont {Y.}~\bibnamefont
  {Lan}}, \bibinfo {author} {\bibfnamefont {D.}~\bibnamefont {Wang}}, \bibinfo
  {author} {\bibfnamefont {G.}~\bibnamefont {Chen}}, \ and\ \bibinfo {author}
  {\bibfnamefont {Z.}~\bibnamefont {Ren}},\ }\href@noop {} {\bibfield
  {journal} {\bibinfo  {journal} {Nano letters}\ }\textbf {\bibinfo {volume}
  {10}},\ \bibinfo {pages} {3373} (\bibinfo {year} {2010})}\BibitemShut
  {NoStop}%
\bibitem [{\citenamefont {Shi}\ \emph {et~al.}(2015)\citenamefont {Shi},
  \citenamefont {Parker}, \citenamefont {Du},\ and\ \citenamefont
  {Singh}}]{shi2015connecting}%
  \BibitemOpen
  \bibfield  {author} {\bibinfo {author} {\bibfnamefont {H.}~\bibnamefont
  {Shi}}, \bibinfo {author} {\bibfnamefont {D.}~\bibnamefont {Parker}},
  \bibinfo {author} {\bibfnamefont {M.-H.}\ \bibnamefont {Du}}, \ and\ \bibinfo
  {author} {\bibfnamefont {D.~J.}\ \bibnamefont {Singh}},\ }\href@noop {}
  {\bibfield  {journal} {\bibinfo  {journal} {Physical Review Applied}\
  }\textbf {\bibinfo {volume} {3}},\ \bibinfo {pages} {014004} (\bibinfo {year}
  {2015})}\BibitemShut {NoStop}%
\bibitem [{\citenamefont {Goldsmid}(1956)}]{goldsmid1956thermal}%
  \BibitemOpen
  \bibfield  {author} {\bibinfo {author} {\bibfnamefont {H.}~\bibnamefont
  {Goldsmid}},\ }\href@noop {} {\bibfield  {journal} {\bibinfo  {journal}
  {Proceedings of the Physical Society. Section B}\ }\textbf {\bibinfo {volume}
  {69}},\ \bibinfo {pages} {203} (\bibinfo {year} {1956})}\BibitemShut
  {NoStop}%
\bibitem [{\citenamefont {Tritt}\ and\ \citenamefont
  {Subramanian}(2006)}]{tritt2006thermoelectric}%
  \BibitemOpen
  \bibfield  {author} {\bibinfo {author} {\bibfnamefont {T.~M.}\ \bibnamefont
  {Tritt}}\ and\ \bibinfo {author} {\bibfnamefont {M.}~\bibnamefont
  {Subramanian}},\ }\href@noop {} {\bibfield  {journal} {\bibinfo  {journal}
  {MRS bulletin}\ }\textbf {\bibinfo {volume} {31}},\ \bibinfo {pages} {188}
  (\bibinfo {year} {2006})}\BibitemShut {NoStop}%
\bibitem [{\citenamefont {Zhang}\ \emph {et~al.}(2011)\citenamefont {Zhang},
  \citenamefont {Kirk}, \citenamefont {Jauregui}, \citenamefont {Yang},
  \citenamefont {Xu}, \citenamefont {Chen},\ and\ \citenamefont
  {Wu}}]{zhang2011rational}%
  \BibitemOpen
  \bibfield  {author} {\bibinfo {author} {\bibfnamefont {G.}~\bibnamefont
  {Zhang}}, \bibinfo {author} {\bibfnamefont {B.}~\bibnamefont {Kirk}},
  \bibinfo {author} {\bibfnamefont {L.~A.}\ \bibnamefont {Jauregui}}, \bibinfo
  {author} {\bibfnamefont {H.}~\bibnamefont {Yang}}, \bibinfo {author}
  {\bibfnamefont {X.}~\bibnamefont {Xu}}, \bibinfo {author} {\bibfnamefont
  {Y.~P.}\ \bibnamefont {Chen}}, \ and\ \bibinfo {author} {\bibfnamefont
  {Y.}~\bibnamefont {Wu}},\ }\href@noop {} {\bibfield  {journal} {\bibinfo
  {journal} {Nano letters}\ }\textbf {\bibinfo {volume} {12}},\ \bibinfo
  {pages} {56} (\bibinfo {year} {2011})}\BibitemShut {NoStop}%
\bibitem [{\citenamefont {Sharma}\ and\ \citenamefont
  {Schwingenschl\"{o}gl}(2016)}]{Sharma_2016}%
  \BibitemOpen
  \bibfield  {author} {\bibinfo {author} {\bibfnamefont {S.}~\bibnamefont
  {Sharma}}\ and\ \bibinfo {author} {\bibfnamefont {U.}~\bibnamefont
  {Schwingenschl\"{o}gl}},\ }\href@noop {} {\bibfield  {journal} {\bibinfo
  {journal} {ACS Energy Lett.}\ }\textbf {\bibinfo {volume} {1}},\ \bibinfo
  {pages} {875} (\bibinfo {year} {2016})}\BibitemShut {NoStop}%
\bibitem [{\citenamefont {Slack}(1973)}]{slack1973nonmetallic}%
  \BibitemOpen
  \bibfield  {author} {\bibinfo {author} {\bibfnamefont {G.~A.}\ \bibnamefont
  {Slack}},\ }\href@noop {} {\bibfield  {journal} {\bibinfo  {journal} {Journal
  of Physics and Chemistry of Solids}\ }\textbf {\bibinfo {volume} {34}},\
  \bibinfo {pages} {321} (\bibinfo {year} {1973})}\BibitemShut {NoStop}%
\bibitem [{\citenamefont {Lindsay}\ \emph {et~al.}(2013)\citenamefont
  {Lindsay}, \citenamefont {Broido},\ and\ \citenamefont
  {Reinecke}}]{lindsay2013first}%
  \BibitemOpen
  \bibfield  {author} {\bibinfo {author} {\bibfnamefont {L.}~\bibnamefont
  {Lindsay}}, \bibinfo {author} {\bibfnamefont {D.}~\bibnamefont {Broido}}, \
  and\ \bibinfo {author} {\bibfnamefont {T.}~\bibnamefont {Reinecke}},\
  }\href@noop {} {\bibfield  {journal} {\bibinfo  {journal} {Physical review
  letters}\ }\textbf {\bibinfo {volume} {111}},\ \bibinfo {pages} {025901}
  (\bibinfo {year} {2013})}\BibitemShut {NoStop}%
\bibitem [{\citenamefont {Naghavi}\ \emph {et~al.}(2018)\citenamefont
  {Naghavi}, \citenamefont {He}, \citenamefont {Xia},\ and\ \citenamefont
  {Wolverton}}]{naghavi2018pd2se3}%
  \BibitemOpen
  \bibfield  {author} {\bibinfo {author} {\bibfnamefont {S.~S.}\ \bibnamefont
  {Naghavi}}, \bibinfo {author} {\bibfnamefont {J.}~\bibnamefont {He}},
  \bibinfo {author} {\bibfnamefont {Y.}~\bibnamefont {Xia}}, \ and\ \bibinfo
  {author} {\bibfnamefont {C.}~\bibnamefont {Wolverton}},\ }\href@noop {}
  {\bibfield  {journal} {\bibinfo  {journal} {Chemistry of Materials}\ }\textbf
  {\bibinfo {volume} {30}},\ \bibinfo {pages} {5639} (\bibinfo {year}
  {2018})}\BibitemShut {NoStop}%
\bibitem [{\citenamefont {Peng}\ \emph
  {et~al.}(2016{\natexlab{a}})\citenamefont {Peng}, \citenamefont {Zhang},
  \citenamefont {Shao}, \citenamefont {Xu}, \citenamefont {Zhang},
  \citenamefont {Lu}, \citenamefont {Zhang},\ and\ \citenamefont
  {Zhu}}]{peng2016first}%
  \BibitemOpen
  \bibfield  {author} {\bibinfo {author} {\bibfnamefont {B.}~\bibnamefont
  {Peng}}, \bibinfo {author} {\bibfnamefont {H.}~\bibnamefont {Zhang}},
  \bibinfo {author} {\bibfnamefont {H.}~\bibnamefont {Shao}}, \bibinfo {author}
  {\bibfnamefont {Y.}~\bibnamefont {Xu}}, \bibinfo {author} {\bibfnamefont
  {R.}~\bibnamefont {Zhang}}, \bibinfo {author} {\bibfnamefont
  {H.}~\bibnamefont {Lu}}, \bibinfo {author} {\bibfnamefont {D.~W.}\
  \bibnamefont {Zhang}}, \ and\ \bibinfo {author} {\bibfnamefont
  {H.}~\bibnamefont {Zhu}},\ }\href@noop {} {\bibfield  {journal} {\bibinfo
  {journal} {ACS applied materials \& interfaces}\ }\textbf {\bibinfo {volume}
  {8}},\ \bibinfo {pages} {20977} (\bibinfo {year}
  {2016}{\natexlab{a}})}\BibitemShut {NoStop}%
\bibitem [{\citenamefont {Ceder}\ and\ \citenamefont
  {Persson}(2010)}]{ceder2010materials}%
  \BibitemOpen
  \bibfield  {author} {\bibinfo {author} {\bibfnamefont {G.}~\bibnamefont
  {Ceder}}\ and\ \bibinfo {author} {\bibfnamefont {K.}~\bibnamefont
  {Persson}},\ }\href@noop {} {\enquote {\bibinfo {title} {The materials
  project: A materials genome approach},}\ } (\bibinfo {year}
  {2010})\BibitemShut {NoStop}%
\bibitem [{\citenamefont {Ong}\ \emph {et~al.}(2008)\citenamefont {Ong},
  \citenamefont {Wang}, \citenamefont {Kang},\ and\ \citenamefont
  {Ceder}}]{ong2008li}%
  \BibitemOpen
  \bibfield  {author} {\bibinfo {author} {\bibfnamefont {S.~P.}\ \bibnamefont
  {Ong}}, \bibinfo {author} {\bibfnamefont {L.}~\bibnamefont {Wang}}, \bibinfo
  {author} {\bibfnamefont {B.}~\bibnamefont {Kang}}, \ and\ \bibinfo {author}
  {\bibfnamefont {G.}~\bibnamefont {Ceder}},\ }\href@noop {} {\bibfield
  {journal} {\bibinfo  {journal} {Chemistry of Materials}\ }\textbf {\bibinfo
  {volume} {20}},\ \bibinfo {pages} {1798} (\bibinfo {year}
  {2008})}\BibitemShut {NoStop}%
\bibitem [{\citenamefont {Jain}\ \emph {et~al.}(2011)\citenamefont {Jain},
  \citenamefont {Hautier}, \citenamefont {Ong}, \citenamefont {Moore},
  \citenamefont {Fischer}, \citenamefont {Persson},\ and\ \citenamefont
  {Ceder}}]{jain2011formation}%
  \BibitemOpen
  \bibfield  {author} {\bibinfo {author} {\bibfnamefont {A.}~\bibnamefont
  {Jain}}, \bibinfo {author} {\bibfnamefont {G.}~\bibnamefont {Hautier}},
  \bibinfo {author} {\bibfnamefont {S.~P.}\ \bibnamefont {Ong}}, \bibinfo
  {author} {\bibfnamefont {C.~J.}\ \bibnamefont {Moore}}, \bibinfo {author}
  {\bibfnamefont {C.~C.}\ \bibnamefont {Fischer}}, \bibinfo {author}
  {\bibfnamefont {K.~A.}\ \bibnamefont {Persson}}, \ and\ \bibinfo {author}
  {\bibfnamefont {G.}~\bibnamefont {Ceder}},\ }\href@noop {} {\bibfield
  {journal} {\bibinfo  {journal} {Physical Review B}\ }\textbf {\bibinfo
  {volume} {84}},\ \bibinfo {pages} {045115} (\bibinfo {year}
  {2011})}\BibitemShut {NoStop}%
\bibitem [{\citenamefont {Mounet}\ \emph {et~al.}(2018)\citenamefont {Mounet},
  \citenamefont {Gibertini}, \citenamefont {Schwaller}, \citenamefont {Campi},
  \citenamefont {Merkys}, \citenamefont {Marrazzo}, \citenamefont {Sohier},
  \citenamefont {Castelli}, \citenamefont {Cepellotti}, \citenamefont {Pizzi}
  \emph {et~al.}}]{mounet2018two}%
  \BibitemOpen
  \bibfield  {author} {\bibinfo {author} {\bibfnamefont {N.}~\bibnamefont
  {Mounet}}, \bibinfo {author} {\bibfnamefont {M.}~\bibnamefont {Gibertini}},
  \bibinfo {author} {\bibfnamefont {P.}~\bibnamefont {Schwaller}}, \bibinfo
  {author} {\bibfnamefont {D.}~\bibnamefont {Campi}}, \bibinfo {author}
  {\bibfnamefont {A.}~\bibnamefont {Merkys}}, \bibinfo {author} {\bibfnamefont
  {A.}~\bibnamefont {Marrazzo}}, \bibinfo {author} {\bibfnamefont
  {T.}~\bibnamefont {Sohier}}, \bibinfo {author} {\bibfnamefont {I.~E.}\
  \bibnamefont {Castelli}}, \bibinfo {author} {\bibfnamefont {A.}~\bibnamefont
  {Cepellotti}}, \bibinfo {author} {\bibfnamefont {G.}~\bibnamefont {Pizzi}},
  \emph {et~al.},\ }\href@noop {} {\bibfield  {journal} {\bibinfo  {journal}
  {Nature nanotechnology}\ }\textbf {\bibinfo {volume} {13}},\ \bibinfo {pages}
  {246} (\bibinfo {year} {2018})}\BibitemShut {NoStop}%
\bibitem [{\citenamefont {Kanagaraj}\ \emph {et~al.}(2019)\citenamefont
  {Kanagaraj}, \citenamefont {Pawbake}, \citenamefont {Sarma}, \citenamefont
  {Rajaji}, \citenamefont {Narayana}, \citenamefont {Measson},\ and\
  \citenamefont {Peter}}]{kanagaraj2019structural}%
  \BibitemOpen
  \bibfield  {author} {\bibinfo {author} {\bibfnamefont {M.}~\bibnamefont
  {Kanagaraj}}, \bibinfo {author} {\bibfnamefont {A.}~\bibnamefont {Pawbake}},
  \bibinfo {author} {\bibfnamefont {S.~C.}\ \bibnamefont {Sarma}}, \bibinfo
  {author} {\bibfnamefont {V.}~\bibnamefont {Rajaji}}, \bibinfo {author}
  {\bibfnamefont {C.}~\bibnamefont {Narayana}}, \bibinfo {author}
  {\bibfnamefont {M.-A.}\ \bibnamefont {Measson}}, \ and\ \bibinfo {author}
  {\bibfnamefont {S.~C.}\ \bibnamefont {Peter}},\ }\href@noop {} {\bibfield
  {journal} {\bibinfo  {journal} {Journal of Alloys and Compounds}\ }\textbf
  {\bibinfo {volume} {794}},\ \bibinfo {pages} {195} (\bibinfo {year}
  {2019})}\BibitemShut {NoStop}%
\bibitem [{\citenamefont {Kresse}\ and\ \citenamefont
  {Hafner}(1993)}]{kresse1993ab}%
  \BibitemOpen
  \bibfield  {author} {\bibinfo {author} {\bibfnamefont {G.}~\bibnamefont
  {Kresse}}\ and\ \bibinfo {author} {\bibfnamefont {J.}~\bibnamefont
  {Hafner}},\ }\href@noop {} {\bibfield  {journal} {\bibinfo  {journal}
  {Physical Review B}\ }\textbf {\bibinfo {volume} {48}},\ \bibinfo {pages}
  {13115} (\bibinfo {year} {1993})}\BibitemShut {NoStop}%
\bibitem [{\citenamefont {Kresse}\ and\ \citenamefont
  {Furthm{\"u}ller}(1996)}]{kresse1996efficiency}%
  \BibitemOpen
  \bibfield  {author} {\bibinfo {author} {\bibfnamefont {G.}~\bibnamefont
  {Kresse}}\ and\ \bibinfo {author} {\bibfnamefont {J.}~\bibnamefont
  {Furthm{\"u}ller}},\ }\href@noop {} {\bibfield  {journal} {\bibinfo
  {journal} {Computational materials science}\ }\textbf {\bibinfo {volume}
  {6}},\ \bibinfo {pages} {15} (\bibinfo {year} {1996})}\BibitemShut {NoStop}%
\bibitem [{\citenamefont {Bl{\"o}chl}(1994)}]{blochl1994projector}%
  \BibitemOpen
  \bibfield  {author} {\bibinfo {author} {\bibfnamefont {P.~E.}\ \bibnamefont
  {Bl{\"o}chl}},\ }\href@noop {} {\bibfield  {journal} {\bibinfo  {journal}
  {Physical review B}\ }\textbf {\bibinfo {volume} {50}},\ \bibinfo {pages}
  {17953} (\bibinfo {year} {1994})}\BibitemShut {NoStop}%
\bibitem [{\citenamefont {Kresse}\ and\ \citenamefont
  {Joubert}(1999)}]{kresse1999ultrasoft}%
  \BibitemOpen
  \bibfield  {author} {\bibinfo {author} {\bibfnamefont {G.}~\bibnamefont
  {Kresse}}\ and\ \bibinfo {author} {\bibfnamefont {D.}~\bibnamefont
  {Joubert}},\ }\href@noop {} {\bibfield  {journal} {\bibinfo  {journal}
  {Physical review b}\ }\textbf {\bibinfo {volume} {59}},\ \bibinfo {pages}
  {1758} (\bibinfo {year} {1999})}\BibitemShut {NoStop}%
\bibitem [{\citenamefont {Perdew}\ \emph {et~al.}(1996)\citenamefont {Perdew},
  \citenamefont {Burke},\ and\ \citenamefont
  {Ernzerhof}}]{perdew1996generalized}%
  \BibitemOpen
  \bibfield  {author} {\bibinfo {author} {\bibfnamefont {J.~P.}\ \bibnamefont
  {Perdew}}, \bibinfo {author} {\bibfnamefont {K.}~\bibnamefont {Burke}}, \
  and\ \bibinfo {author} {\bibfnamefont {M.}~\bibnamefont {Ernzerhof}},\
  }\href@noop {} {\bibfield  {journal} {\bibinfo  {journal} {Physical review
  letters}\ }\textbf {\bibinfo {volume} {77}},\ \bibinfo {pages} {3865}
  (\bibinfo {year} {1996})}\BibitemShut {NoStop}%
\bibitem [{\citenamefont {Madsen}\ and\ \citenamefont
  {Singh}(2006)}]{madsen2006boltztrap}%
  \BibitemOpen
  \bibfield  {author} {\bibinfo {author} {\bibfnamefont {G.~K.}\ \bibnamefont
  {Madsen}}\ and\ \bibinfo {author} {\bibfnamefont {D.~J.}\ \bibnamefont
  {Singh}},\ }\href@noop {} {\bibfield  {journal} {\bibinfo  {journal}
  {Computer Physics Communications}\ }\textbf {\bibinfo {volume} {175}},\
  \bibinfo {pages} {67} (\bibinfo {year} {2006})}\BibitemShut {NoStop}%
\bibitem [{\citenamefont {Togo}\ \emph {et~al.}(2008)\citenamefont {Togo},
  \citenamefont {Oba},\ and\ \citenamefont {Tanaka}}]{togo2008first}%
  \BibitemOpen
  \bibfield  {author} {\bibinfo {author} {\bibfnamefont {A.}~\bibnamefont
  {Togo}}, \bibinfo {author} {\bibfnamefont {F.}~\bibnamefont {Oba}}, \ and\
  \bibinfo {author} {\bibfnamefont {I.}~\bibnamefont {Tanaka}},\ }\href@noop {}
  {\bibfield  {journal} {\bibinfo  {journal} {Physical Review B}\ }\textbf
  {\bibinfo {volume} {78}},\ \bibinfo {pages} {134106} (\bibinfo {year}
  {2008})}\BibitemShut {NoStop}%
\bibitem [{\citenamefont {Li}\ \emph {et~al.}(2014)\citenamefont {Li},
  \citenamefont {Carrete}, \citenamefont {Katcho},\ and\ \citenamefont
  {Mingo}}]{li2014shengbte}%
  \BibitemOpen
  \bibfield  {author} {\bibinfo {author} {\bibfnamefont {W.}~\bibnamefont
  {Li}}, \bibinfo {author} {\bibfnamefont {J.}~\bibnamefont {Carrete}},
  \bibinfo {author} {\bibfnamefont {N.~A.}\ \bibnamefont {Katcho}}, \ and\
  \bibinfo {author} {\bibfnamefont {N.}~\bibnamefont {Mingo}},\ }\href@noop {}
  {\bibfield  {journal} {\bibinfo  {journal} {Computer Physics Communications}\
  }\textbf {\bibinfo {volume} {185}},\ \bibinfo {pages} {1747} (\bibinfo {year}
  {2014})}\BibitemShut {NoStop}%
\bibitem [{\citenamefont {Luo}\ \emph {et~al.}(2012)\citenamefont {Luo},
  \citenamefont {Sullivan},\ and\ \citenamefont {Quek}}]{luo2012first}%
  \BibitemOpen
  \bibfield  {author} {\bibinfo {author} {\bibfnamefont {X.}~\bibnamefont
  {Luo}}, \bibinfo {author} {\bibfnamefont {M.~B.}\ \bibnamefont {Sullivan}}, \
  and\ \bibinfo {author} {\bibfnamefont {S.~Y.}\ \bibnamefont {Quek}},\
  }\href@noop {} {\bibfield  {journal} {\bibinfo  {journal} {Physical Review
  B}\ }\textbf {\bibinfo {volume} {86}},\ \bibinfo {pages} {184111} (\bibinfo
  {year} {2012})}\BibitemShut {NoStop}%
\bibitem [{\citenamefont {Bj{\"o}rkman}\ \emph {et~al.}(2012)\citenamefont
  {Bj{\"o}rkman}, \citenamefont {Gulans}, \citenamefont {Krasheninnikov},\ and\
  \citenamefont {Nieminen}}]{bjorkman2012van}%
  \BibitemOpen
  \bibfield  {author} {\bibinfo {author} {\bibfnamefont {T.}~\bibnamefont
  {Bj{\"o}rkman}}, \bibinfo {author} {\bibfnamefont {A.}~\bibnamefont
  {Gulans}}, \bibinfo {author} {\bibfnamefont {A.~V.}\ \bibnamefont
  {Krasheninnikov}}, \ and\ \bibinfo {author} {\bibfnamefont {R.~M.}\
  \bibnamefont {Nieminen}},\ }\href@noop {} {\bibfield  {journal} {\bibinfo
  {journal} {Physical review letters}\ }\textbf {\bibinfo {volume} {108}},\
  \bibinfo {pages} {235502} (\bibinfo {year} {2012})}\BibitemShut {NoStop}%
\bibitem [{\citenamefont {Ambrosi}\ and\ \citenamefont
  {Pumera}(2018)}]{ambrosi2018exfoliation}%
  \BibitemOpen
  \bibfield  {author} {\bibinfo {author} {\bibfnamefont {A.}~\bibnamefont
  {Ambrosi}}\ and\ \bibinfo {author} {\bibfnamefont {M.}~\bibnamefont
  {Pumera}},\ }\href@noop {} {\bibfield  {journal} {\bibinfo  {journal}
  {Chemical Society Reviews}\ }\textbf {\bibinfo {volume} {47}},\ \bibinfo
  {pages} {7213} (\bibinfo {year} {2018})}\BibitemShut {NoStop}%
\bibitem [{\citenamefont {Greenaway}\ and\ \citenamefont
  {Harbeke}(1965)}]{greenaway1965band}%
  \BibitemOpen
  \bibfield  {author} {\bibinfo {author} {\bibfnamefont {D.~L.}\ \bibnamefont
  {Greenaway}}\ and\ \bibinfo {author} {\bibfnamefont {G.}~\bibnamefont
  {Harbeke}},\ }\href@noop {} {\bibfield  {journal} {\bibinfo  {journal}
  {Journal of Physics and Chemistry of Solids}\ }\textbf {\bibinfo {volume}
  {26}},\ \bibinfo {pages} {1585} (\bibinfo {year} {1965})}\BibitemShut
  {NoStop}%
\bibitem [{\citenamefont {Becke}\ and\ \citenamefont {Edgecombe}(1990)}]{ELF1}%
  \BibitemOpen
  \bibfield  {author} {\bibinfo {author} {\bibfnamefont {A.~D.}\ \bibnamefont
  {Becke}}\ and\ \bibinfo {author} {\bibfnamefont {K.~E.}\ \bibnamefont
  {Edgecombe}},\ }\href@noop {} {\bibfield  {journal} {\bibinfo  {journal} {The
  Journal of chemical physics}\ }\textbf {\bibinfo {volume} {92}},\ \bibinfo
  {pages} {5397} (\bibinfo {year} {1990})}\BibitemShut {NoStop}%
\bibitem [{\citenamefont {Walsh}\ \emph {et~al.}(2011)\citenamefont {Walsh},
  \citenamefont {Payne}, \citenamefont {Egdell},\ and\ \citenamefont
  {Watson}}]{Walsh2011}%
  \BibitemOpen
  \bibfield  {author} {\bibinfo {author} {\bibfnamefont {A.}~\bibnamefont
  {Walsh}}, \bibinfo {author} {\bibfnamefont {D.~J.}\ \bibnamefont {Payne}},
  \bibinfo {author} {\bibfnamefont {R.~G.}\ \bibnamefont {Egdell}}, \ and\
  \bibinfo {author} {\bibfnamefont {G.~W.}\ \bibnamefont {Watson}},\
  }\href@noop {} {\bibfield  {journal} {\bibinfo  {journal} {Chemical Society
  Reviews}\ }\textbf {\bibinfo {volume} {40}},\ \bibinfo {pages} {4455}
  (\bibinfo {year} {2011})}\BibitemShut {NoStop}%
\bibitem [{\citenamefont {Witting}\ \emph {et~al.}(2019)\citenamefont
  {Witting}, \citenamefont {Chasapis}, \citenamefont {Ricci}, \citenamefont
  {Peters}, \citenamefont {Heinz}, \citenamefont {Hautier},\ and\ \citenamefont
  {Snyder}}]{Witting2019}%
  \BibitemOpen
  \bibfield  {author} {\bibinfo {author} {\bibfnamefont {I.~T.}\ \bibnamefont
  {Witting}}, \bibinfo {author} {\bibfnamefont {T.~C.}\ \bibnamefont
  {Chasapis}}, \bibinfo {author} {\bibfnamefont {F.}~\bibnamefont {Ricci}},
  \bibinfo {author} {\bibfnamefont {M.}~\bibnamefont {Peters}}, \bibinfo
  {author} {\bibfnamefont {N.~A.}\ \bibnamefont {Heinz}}, \bibinfo {author}
  {\bibfnamefont {G.}~\bibnamefont {Hautier}}, \ and\ \bibinfo {author}
  {\bibfnamefont {G.~J.}\ \bibnamefont {Snyder}},\ }\href@noop {} {\bibfield
  {journal} {\bibinfo  {journal} {Advanced Electronic Materials}\ ,\ \bibinfo
  {pages} {1800904}} (\bibinfo {year} {2019})}\BibitemShut {NoStop}%
\bibitem [{\citenamefont {Lee}\ and\ \citenamefont {Elliott}(2017)}]{Lee2017}%
  \BibitemOpen
  \bibfield  {author} {\bibinfo {author} {\bibfnamefont {T.~H.}\ \bibnamefont
  {Lee}}\ and\ \bibinfo {author} {\bibfnamefont {S.~R.}\ \bibnamefont
  {Elliott}},\ }\href@noop {} {\bibfield  {journal} {\bibinfo  {journal}
  {Advanced Materials}\ }\textbf {\bibinfo {volume} {29}},\ \bibinfo {pages}
  {1700814} (\bibinfo {year} {2017})}\BibitemShut {NoStop}%
\bibitem [{\citenamefont {Da~Silva}\ \emph {et~al.}(2008)\citenamefont
  {Da~Silva}, \citenamefont {Walsh},\ and\ \citenamefont {Lee}}]{DaSilva2008}%
  \BibitemOpen
  \bibfield  {author} {\bibinfo {author} {\bibfnamefont {J.~L.}\ \bibnamefont
  {Da~Silva}}, \bibinfo {author} {\bibfnamefont {A.}~\bibnamefont {Walsh}}, \
  and\ \bibinfo {author} {\bibfnamefont {H.}~\bibnamefont {Lee}},\ }\href@noop
  {} {\bibfield  {journal} {\bibinfo  {journal} {Physical Review B}\ }\textbf
  {\bibinfo {volume} {78}},\ \bibinfo {pages} {224111} (\bibinfo {year}
  {2008})}\BibitemShut {NoStop}%
\bibitem [{\citenamefont {Huang}\ and\ \citenamefont
  {Zeng}(2015)}]{huang2015roles}%
  \BibitemOpen
  \bibfield  {author} {\bibinfo {author} {\bibfnamefont {L.-F.}\ \bibnamefont
  {Huang}}\ and\ \bibinfo {author} {\bibfnamefont {Z.}~\bibnamefont {Zeng}},\
  }\href@noop {} {\bibfield  {journal} {\bibinfo  {journal} {The Journal of
  Physical Chemistry C}\ }\textbf {\bibinfo {volume} {119}},\ \bibinfo {pages}
  {18779} (\bibinfo {year} {2015})}\BibitemShut {NoStop}%
\bibitem [{\citenamefont {Deringer}\ \emph {et~al.}(2011)\citenamefont
  {Deringer}, \citenamefont {Tchougréeff},\ and\ \citenamefont
  {Dronskowski}}]{Deringer_2011}%
  \BibitemOpen
  \bibfield  {author} {\bibinfo {author} {\bibfnamefont {V.~L.}\ \bibnamefont
  {Deringer}}, \bibinfo {author} {\bibfnamefont {A.~L.}\ \bibnamefont
  {Tchougréeff}}, \ and\ \bibinfo {author} {\bibfnamefont {R.}~\bibnamefont
  {Dronskowski}},\ }\href@noop {} {\bibfield  {journal} {\bibinfo  {journal}
  {J. Phys. Chem.\,A}\ }\textbf {\bibinfo {volume} {115}},\ \bibinfo {pages}
  {5461} (\bibinfo {year} {2011})}\BibitemShut {NoStop}%
\bibitem [{\citenamefont {Dronskowski}\ and\ \citenamefont
  {Bloechl}(1993)}]{Dronskowski_1993}%
  \BibitemOpen
  \bibfield  {author} {\bibinfo {author} {\bibfnamefont {R.}~\bibnamefont
  {Dronskowski}}\ and\ \bibinfo {author} {\bibfnamefont {P.~E.}\ \bibnamefont
  {Bloechl}},\ }\href@noop {} {\bibfield  {journal} {\bibinfo  {journal} {J.
  Phys. Chem.}\ }\textbf {\bibinfo {volume} {97}},\ \bibinfo {pages} {8617}
  (\bibinfo {year} {1993})}\BibitemShut {NoStop}%
\bibitem [{\citenamefont {Hoffmann}(1987)}]{Hoffmann_1987}%
  \BibitemOpen
  \bibfield  {author} {\bibinfo {author} {\bibfnamefont {R.}~\bibnamefont
  {Hoffmann}},\ }\href@noop {} {\bibfield  {journal} {\bibinfo  {journal}
  {Angew. Chem. Int. Ed. Engl.}\ }\textbf {\bibinfo {volume} {26}},\ \bibinfo
  {pages} {846} (\bibinfo {year} {1987})}\BibitemShut {NoStop}%
\bibitem [{\citenamefont {Maintz}\ \emph {et~al.}(2016)\citenamefont {Maintz},
  \citenamefont {Deringer}, \citenamefont {Tchougr{\'e}eff},\ and\
  \citenamefont {Dronskowski}}]{maintz2016lobster}%
  \BibitemOpen
  \bibfield  {author} {\bibinfo {author} {\bibfnamefont {S.}~\bibnamefont
  {Maintz}}, \bibinfo {author} {\bibfnamefont {V.~L.}\ \bibnamefont
  {Deringer}}, \bibinfo {author} {\bibfnamefont {A.~L.}\ \bibnamefont
  {Tchougr{\'e}eff}}, \ and\ \bibinfo {author} {\bibfnamefont {R.}~\bibnamefont
  {Dronskowski}},\ }\href@noop {} {\bibfield  {journal} {\bibinfo  {journal}
  {Journal of computational chemistry}\ }\textbf {\bibinfo {volume} {37}},\
  \bibinfo {pages} {1030} (\bibinfo {year} {2016})}\BibitemShut {NoStop}%
\bibitem [{\citenamefont {Maintz}\ \emph {et~al.}(2013)\citenamefont {Maintz},
  \citenamefont {Deringer}, \citenamefont {Tchougr{\'e}eff},\ and\
  \citenamefont {Dronskowski}}]{maintz2013analytic}%
  \BibitemOpen
  \bibfield  {author} {\bibinfo {author} {\bibfnamefont {S.}~\bibnamefont
  {Maintz}}, \bibinfo {author} {\bibfnamefont {V.~L.}\ \bibnamefont
  {Deringer}}, \bibinfo {author} {\bibfnamefont {A.~L.}\ \bibnamefont
  {Tchougr{\'e}eff}}, \ and\ \bibinfo {author} {\bibfnamefont {R.}~\bibnamefont
  {Dronskowski}},\ }\href@noop {} {\bibfield  {journal} {\bibinfo  {journal}
  {Journal of computational chemistry}\ }\textbf {\bibinfo {volume} {34}},\
  \bibinfo {pages} {2557} (\bibinfo {year} {2013})}\BibitemShut {NoStop}%
\bibitem [{\citenamefont {Kumar}\ and\ \citenamefont
  {Schwingenschlogl}(2015)}]{kumar2015thermoelectric}%
  \BibitemOpen
  \bibfield  {author} {\bibinfo {author} {\bibfnamefont {S.}~\bibnamefont
  {Kumar}}\ and\ \bibinfo {author} {\bibfnamefont {U.}~\bibnamefont
  {Schwingenschlogl}},\ }\href@noop {} {\bibfield  {journal} {\bibinfo
  {journal} {Chemistry of Materials}\ }\textbf {\bibinfo {volume} {27}},\
  \bibinfo {pages} {1278} (\bibinfo {year} {2015})}\BibitemShut {NoStop}%
\bibitem [{\citenamefont {Peng}\ \emph
  {et~al.}(2016{\natexlab{b}})\citenamefont {Peng}, \citenamefont {Zhang},
  \citenamefont {Shao}, \citenamefont {Xu}, \citenamefont {Zhang},\ and\
  \citenamefont {Zhu}}]{Peng_2016}%
  \BibitemOpen
  \bibfield  {author} {\bibinfo {author} {\bibfnamefont {B.}~\bibnamefont
  {Peng}}, \bibinfo {author} {\bibfnamefont {H.}~\bibnamefont {Zhang}},
  \bibinfo {author} {\bibfnamefont {H.}~\bibnamefont {Shao}}, \bibinfo {author}
  {\bibfnamefont {Y.}~\bibnamefont {Xu}}, \bibinfo {author} {\bibfnamefont
  {X.}~\bibnamefont {Zhang}}, \ and\ \bibinfo {author} {\bibfnamefont
  {H.}~\bibnamefont {Zhu}},\ }\href@noop {} {\bibfield  {journal} {\bibinfo
  {journal} {RSC Adv.}\ }\textbf {\bibinfo {volume} {6}},\ \bibinfo {pages}
  {5767} (\bibinfo {year} {2016}{\natexlab{b}})}\BibitemShut {NoStop}%
\bibitem [{\citenamefont {Gandi}\ and\ \citenamefont
  {Schwingenschl{\"{o}}gl}(2016)}]{Gandi2016}%
  \BibitemOpen
  \bibfield  {author} {\bibinfo {author} {\bibfnamefont {A.~N.}\ \bibnamefont
  {Gandi}}\ and\ \bibinfo {author} {\bibfnamefont {U.}~\bibnamefont
  {Schwingenschl{\"{o}}gl}},\ }\href@noop {} {\bibfield  {journal} {\bibinfo
  {journal} {EPL (Europhysics Lett.}\ }\textbf {\bibinfo {volume} {113}},\
  \bibinfo {pages} {36002} (\bibinfo {year} {2016})}\BibitemShut {NoStop}%
\bibitem [{\citenamefont {Dutta}\ \emph {et~al.}(2019)\citenamefont {Dutta},
  \citenamefont {Pal}, \citenamefont {Waghmare},\ and\ \citenamefont
  {Biswas}}]{Dutta2019}%
  \BibitemOpen
  \bibfield  {author} {\bibinfo {author} {\bibfnamefont {M.}~\bibnamefont
  {Dutta}}, \bibinfo {author} {\bibfnamefont {K.}~\bibnamefont {Pal}}, \bibinfo
  {author} {\bibfnamefont {U.~V.}\ \bibnamefont {Waghmare}}, \ and\ \bibinfo
  {author} {\bibfnamefont {K.}~\bibnamefont {Biswas}},\ }\href@noop {}
  {\bibfield  {journal} {\bibinfo  {journal} {Chemical science}\ }\textbf
  {\bibinfo {volume} {10}},\ \bibinfo {pages} {4905} (\bibinfo {year}
  {2019})}\BibitemShut {NoStop}%
\bibitem [{\citenamefont {Nielsen}\ \emph {et~al.}(2013)\citenamefont
  {Nielsen}, \citenamefont {Ozolins},\ and\ \citenamefont
  {Heremans}}]{Nielsen2013}%
  \BibitemOpen
  \bibfield  {author} {\bibinfo {author} {\bibfnamefont {M.~D.}\ \bibnamefont
  {Nielsen}}, \bibinfo {author} {\bibfnamefont {V.}~\bibnamefont {Ozolins}}, \
  and\ \bibinfo {author} {\bibfnamefont {J.~P.}\ \bibnamefont {Heremans}},\
  }\href@noop {} {\bibfield  {journal} {\bibinfo  {journal} {Energy \&
  Environmental Science}\ }\textbf {\bibinfo {volume} {6}},\ \bibinfo {pages}
  {570} (\bibinfo {year} {2013})}\BibitemShut {NoStop}%
\bibitem [{\citenamefont {Deringer}\ \emph {et~al.}(2015)\citenamefont
  {Deringer}, \citenamefont {Stoffel}, \citenamefont {Wuttig},\ and\
  \citenamefont {Dronskowski}}]{Deringer_2015}%
  \BibitemOpen
  \bibfield  {author} {\bibinfo {author} {\bibfnamefont {V.~L.}\ \bibnamefont
  {Deringer}}, \bibinfo {author} {\bibfnamefont {R.~P.}\ \bibnamefont
  {Stoffel}}, \bibinfo {author} {\bibfnamefont {M.}~\bibnamefont {Wuttig}}, \
  and\ \bibinfo {author} {\bibfnamefont {R.}~\bibnamefont {Dronskowski}},\
  }\href@noop {} {\bibfield  {journal} {\bibinfo  {journal} {Chemical Science}\
  }\textbf {\bibinfo {volume} {6}},\ \bibinfo {pages} {5255} (\bibinfo {year}
  {2015})}\BibitemShut {NoStop}%
\bibitem [{\citenamefont {Hung}\ \emph {et~al.}(2019)\citenamefont {Hung},
  \citenamefont {Nugraha},\ and\ \citenamefont {Saito}}]{hung2019designing}%
  \BibitemOpen
  \bibfield  {author} {\bibinfo {author} {\bibfnamefont {N.~T.}\ \bibnamefont
  {Hung}}, \bibinfo {author} {\bibfnamefont {A.~R.}\ \bibnamefont {Nugraha}}, \
  and\ \bibinfo {author} {\bibfnamefont {R.}~\bibnamefont {Saito}},\
  }\href@noop {} {\bibfield  {journal} {\bibinfo  {journal} {Nano Energy}\
  }\textbf {\bibinfo {volume} {58}},\ \bibinfo {pages} {743} (\bibinfo {year}
  {2019})}\BibitemShut {NoStop}%
\bibitem [{\citenamefont {Lee}\ \emph {et~al.}(2018)\citenamefont {Lee},
  \citenamefont {Kim}, \citenamefont {Tak}, \citenamefont {Cho}, \citenamefont
  {Shim}, \citenamefont {Lim},\ and\ \citenamefont {Whangbo}}]{Lee_2018}%
  \BibitemOpen
  \bibfield  {author} {\bibinfo {author} {\bibfnamefont {C.}~\bibnamefont
  {Lee}}, \bibinfo {author} {\bibfnamefont {J.~N.}\ \bibnamefont {Kim}},
  \bibinfo {author} {\bibfnamefont {J.-Y.}\ \bibnamefont {Tak}}, \bibinfo
  {author} {\bibfnamefont {H.~K.}\ \bibnamefont {Cho}}, \bibinfo {author}
  {\bibfnamefont {J.~H.}\ \bibnamefont {Shim}}, \bibinfo {author}
  {\bibfnamefont {Y.~S.}\ \bibnamefont {Lim}}, \ and\ \bibinfo {author}
  {\bibfnamefont {M.-H.}\ \bibnamefont {Whangbo}},\ }\href@noop {} {\bibfield
  {journal} {\bibinfo  {journal} {AIP Adv.}\ }\textbf {\bibinfo {volume} {8}},\
  \bibinfo {pages} {115213} (\bibinfo {year} {2018})}\BibitemShut {NoStop}%
\bibitem [{\citenamefont {Zhou}\ and\ \citenamefont {Wang}(2015)}]{Zhou_2015}%
  \BibitemOpen
  \bibfield  {author} {\bibinfo {author} {\bibfnamefont {G.}~\bibnamefont
  {Zhou}}\ and\ \bibinfo {author} {\bibfnamefont {D.}~\bibnamefont {Wang}},\
  }\href@noop {} {\bibfield  {journal} {\bibinfo  {journal} {Sci. Rep.}\
  }\textbf {\bibinfo {volume} {5}} (\bibinfo {year} {2015})}\BibitemShut
  {NoStop}%
\bibitem [{\citenamefont {Xu}\ \emph {et~al.}(2018)\citenamefont {Xu},
  \citenamefont {Zhang}, \citenamefont {Yu}, \citenamefont {Ma}, \citenamefont
  {Wang},\ and\ \citenamefont {Wang}}]{Xu_2018}%
  \BibitemOpen
  \bibfield  {author} {\bibinfo {author} {\bibfnamefont {B.}~\bibnamefont
  {Xu}}, \bibinfo {author} {\bibfnamefont {J.}~\bibnamefont {Zhang}}, \bibinfo
  {author} {\bibfnamefont {G.}~\bibnamefont {Yu}}, \bibinfo {author}
  {\bibfnamefont {S.}~\bibnamefont {Ma}}, \bibinfo {author} {\bibfnamefont
  {Y.}~\bibnamefont {Wang}}, \ and\ \bibinfo {author} {\bibfnamefont
  {Y.}~\bibnamefont {Wang}},\ }\href@noop {} {\bibfield  {journal} {\bibinfo
  {journal} {J. Appl. Phys.}\ }\textbf {\bibinfo {volume} {124}},\ \bibinfo
  {pages} {165104} (\bibinfo {year} {2018})}\BibitemShut {NoStop}%
\bibitem [{\citenamefont {Garrity}(2016)}]{Garrity2016}%
  \BibitemOpen
  \bibfield  {author} {\bibinfo {author} {\bibfnamefont {K.~F.}\ \bibnamefont
  {Garrity}},\ }\href@noop {} {\bibfield  {journal} {\bibinfo  {journal}
  {Physical Review B}\ }\textbf {\bibinfo {volume} {94}},\ \bibinfo {pages}
  {045122} (\bibinfo {year} {2016})}\BibitemShut {NoStop}%
\bibitem [{\citenamefont {T.~Hung}\ \emph {et~al.}(2019)\citenamefont
  {T.~Hung}, \citenamefont {Nugraha}, \citenamefont {Yang}, \citenamefont
  {Zhang},\ and\ \citenamefont {Saito}}]{THung2019}%
  \BibitemOpen
  \bibfield  {author} {\bibinfo {author} {\bibfnamefont {N.}~\bibnamefont
  {T.~Hung}}, \bibinfo {author} {\bibfnamefont {A.~R.}\ \bibnamefont
  {Nugraha}}, \bibinfo {author} {\bibfnamefont {T.}~\bibnamefont {Yang}},
  \bibinfo {author} {\bibfnamefont {Z.}~\bibnamefont {Zhang}}, \ and\ \bibinfo
  {author} {\bibfnamefont {R.}~\bibnamefont {Saito}},\ }\href@noop {}
  {\bibfield  {journal} {\bibinfo  {journal} {Journal of Applied Physics}\
  }\textbf {\bibinfo {volume} {125}},\ \bibinfo {pages} {082502} (\bibinfo
  {year} {2019})}\BibitemShut {NoStop}%
\bibitem [{\citenamefont {Hinsche}\ \emph {et~al.}(2011)\citenamefont
  {Hinsche}, \citenamefont {Yavorsky}, \citenamefont {Mertig},\ and\
  \citenamefont {Zahn}}]{Hinsche2011}%
  \BibitemOpen
  \bibfield  {author} {\bibinfo {author} {\bibfnamefont {N.}~\bibnamefont
  {Hinsche}}, \bibinfo {author} {\bibfnamefont {B.~Y.}\ \bibnamefont
  {Yavorsky}}, \bibinfo {author} {\bibfnamefont {I.}~\bibnamefont {Mertig}}, \
  and\ \bibinfo {author} {\bibfnamefont {P.}~\bibnamefont {Zahn}},\ }\href@noop
  {} {\bibfield  {journal} {\bibinfo  {journal} {Physical Review B}\ }\textbf
  {\bibinfo {volume} {84}},\ \bibinfo {pages} {165214} (\bibinfo {year}
  {2011})}\BibitemShut {NoStop}%
\bibitem [{\citenamefont {Scheidemantel}\ \emph {et~al.}(2003)\citenamefont
  {Scheidemantel}, \citenamefont {Ambrosch-Draxl}, \citenamefont {Thonhauser},
  \citenamefont {Badding},\ and\ \citenamefont {Sofo}}]{Scheidemantel2003}%
  \BibitemOpen
  \bibfield  {author} {\bibinfo {author} {\bibfnamefont {T.~J.}\ \bibnamefont
  {Scheidemantel}}, \bibinfo {author} {\bibfnamefont {C.}~\bibnamefont
  {Ambrosch-Draxl}}, \bibinfo {author} {\bibfnamefont {T.}~\bibnamefont
  {Thonhauser}}, \bibinfo {author} {\bibfnamefont {J.~V.}\ \bibnamefont
  {Badding}}, \ and\ \bibinfo {author} {\bibfnamefont {J.~O.}\ \bibnamefont
  {Sofo}},\ }\href@noop {} {\bibfield  {journal} {\bibinfo  {journal} {Phys.
  Rev. B}\ }\textbf {\bibinfo {volume} {68}},\ \bibinfo {pages} {125210}
  (\bibinfo {year} {2003})}\BibitemShut {NoStop}%
\bibitem [{\citenamefont {Joo}\ \emph {et~al.}(2019)\citenamefont {Joo},
  \citenamefont {Ryu}, \citenamefont {Son}, \citenamefont {Lee}, \citenamefont
  {Min},\ and\ \citenamefont {Kim}}]{Joo2019}%
  \BibitemOpen
  \bibfield  {author} {\bibinfo {author} {\bibfnamefont {S.-J.}\ \bibnamefont
  {Joo}}, \bibinfo {author} {\bibfnamefont {B.}~\bibnamefont {Ryu}}, \bibinfo
  {author} {\bibfnamefont {J.-H.}\ \bibnamefont {Son}}, \bibinfo {author}
  {\bibfnamefont {J.~E.}\ \bibnamefont {Lee}}, \bibinfo {author} {\bibfnamefont
  {B.-K.}\ \bibnamefont {Min}}, \ and\ \bibinfo {author} {\bibfnamefont
  {B.-S.}\ \bibnamefont {Kim}},\ }\href@noop {} {\bibfield  {journal} {\bibinfo
   {journal} {Journal of Alloys and Compounds}\ }\textbf {\bibinfo {volume}
  {783}},\ \bibinfo {pages} {448} (\bibinfo {year} {2019})}\BibitemShut
  {NoStop}%
\bibitem [{\citenamefont {Hicks}\ and\ \citenamefont
  {Dresselhaus}(1993)}]{Hicks1993}%
  \BibitemOpen
  \bibfield  {author} {\bibinfo {author} {\bibfnamefont {L.}~\bibnamefont
  {Hicks}}\ and\ \bibinfo {author} {\bibfnamefont {M.~S.}\ \bibnamefont
  {Dresselhaus}},\ }\href@noop {} {\bibfield  {journal} {\bibinfo  {journal}
  {Physical Review B}\ }\textbf {\bibinfo {volume} {47}},\ \bibinfo {pages}
  {12727} (\bibinfo {year} {1993})}\BibitemShut {NoStop}%
\bibitem [{\citenamefont {Babaei}\ \emph {et~al.}(2014)\citenamefont {Babaei},
  \citenamefont {Khodadadi},\ and\ \citenamefont {Sinha}}]{babaei2014large}%
  \BibitemOpen
  \bibfield  {author} {\bibinfo {author} {\bibfnamefont {H.}~\bibnamefont
  {Babaei}}, \bibinfo {author} {\bibfnamefont {J.}~\bibnamefont {Khodadadi}}, \
  and\ \bibinfo {author} {\bibfnamefont {S.}~\bibnamefont {Sinha}},\
  }\href@noop {} {\bibfield  {journal} {\bibinfo  {journal} {Applied Physics
  Letters}\ }\textbf {\bibinfo {volume} {105}},\ \bibinfo {pages} {193901}
  (\bibinfo {year} {2014})}\BibitemShut {NoStop}%
\bibitem [{\citenamefont {Zhang}\ \emph {et~al.}(2017)\citenamefont {Zhang},
  \citenamefont {Liu}, \citenamefont {Wen}, \citenamefont {Shi}, \citenamefont
  {Chen}, \citenamefont {Liu},\ and\ \citenamefont {Shan}}]{Zhang2017}%
  \BibitemOpen
  \bibfield  {author} {\bibinfo {author} {\bibfnamefont {J.}~\bibnamefont
  {Zhang}}, \bibinfo {author} {\bibfnamefont {X.}~\bibnamefont {Liu}}, \bibinfo
  {author} {\bibfnamefont {Y.}~\bibnamefont {Wen}}, \bibinfo {author}
  {\bibfnamefont {L.}~\bibnamefont {Shi}}, \bibinfo {author} {\bibfnamefont
  {R.}~\bibnamefont {Chen}}, \bibinfo {author} {\bibfnamefont {H.}~\bibnamefont
  {Liu}}, \ and\ \bibinfo {author} {\bibfnamefont {B.}~\bibnamefont {Shan}},\
  }\href@noop {} {\bibfield  {journal} {\bibinfo  {journal} {ACS Appl. Mater.
  Interfaces}\ }\textbf {\bibinfo {volume} {9}},\ \bibinfo {pages} {2509}
  (\bibinfo {year} {2017})}\BibitemShut {NoStop}%
\bibitem [{\citenamefont {Wang}\ \emph {et~al.}(2015)\citenamefont {Wang},
  \citenamefont {Zhang}, \citenamefont {Yu},\ and\ \citenamefont
  {Wang}}]{Wang2015}%
  \BibitemOpen
  \bibfield  {author} {\bibinfo {author} {\bibfnamefont {F.~Q.}\ \bibnamefont
  {Wang}}, \bibinfo {author} {\bibfnamefont {S.}~\bibnamefont {Zhang}},
  \bibinfo {author} {\bibfnamefont {J.}~\bibnamefont {Yu}}, \ and\ \bibinfo
  {author} {\bibfnamefont {Q.}~\bibnamefont {Wang}},\ }\href@noop {} {\bibfield
   {journal} {\bibinfo  {journal} {Nanoscale}\ }\textbf {\bibinfo {volume}
  {7}},\ \bibinfo {pages} {15962} (\bibinfo {year} {2015})}\BibitemShut
  {NoStop}%
\bibitem [{\citenamefont {Zhao}\ \emph {et~al.}(2014)\citenamefont {Zhao},
  \citenamefont {Lo}, \citenamefont {Zhang}, \citenamefont {Sun}, \citenamefont
  {Tan}, \citenamefont {Uher}, \citenamefont {Wolverton}, \citenamefont
  {Dravid},\ and\ \citenamefont {Kanatzidis}}]{Zhao2014}%
  \BibitemOpen
  \bibfield  {author} {\bibinfo {author} {\bibfnamefont {L.-D.}\ \bibnamefont
  {Zhao}}, \bibinfo {author} {\bibfnamefont {S.-H.}\ \bibnamefont {Lo}},
  \bibinfo {author} {\bibfnamefont {Y.}~\bibnamefont {Zhang}}, \bibinfo
  {author} {\bibfnamefont {H.}~\bibnamefont {Sun}}, \bibinfo {author}
  {\bibfnamefont {G.}~\bibnamefont {Tan}}, \bibinfo {author} {\bibfnamefont
  {C.}~\bibnamefont {Uher}}, \bibinfo {author} {\bibfnamefont {C.}~\bibnamefont
  {Wolverton}}, \bibinfo {author} {\bibfnamefont {V.~P.}\ \bibnamefont
  {Dravid}}, \ and\ \bibinfo {author} {\bibfnamefont {M.~G.}\ \bibnamefont
  {Kanatzidis}},\ }\href@noop {} {\bibfield  {journal} {\bibinfo  {journal}
  {Nature}\ }\textbf {\bibinfo {volume} {508}},\ \bibinfo {pages} {373}
  (\bibinfo {year} {2014})}\BibitemShut {NoStop}%
\bibitem [{\citenamefont {Yan}\ \emph {et~al.}(2014)\citenamefont {Yan},
  \citenamefont {Simpson}, \citenamefont {Bertolazzi}, \citenamefont {Brivio},
  \citenamefont {Watson}, \citenamefont {Wu}, \citenamefont {Kis},
  \citenamefont {Luo}, \citenamefont {{Hight Walker}},\ and\ \citenamefont
  {Xing}}]{Yan2014}%
  \BibitemOpen
  \bibfield  {author} {\bibinfo {author} {\bibfnamefont {R.}~\bibnamefont
  {Yan}}, \bibinfo {author} {\bibfnamefont {J.~R.}\ \bibnamefont {Simpson}},
  \bibinfo {author} {\bibfnamefont {S.}~\bibnamefont {Bertolazzi}}, \bibinfo
  {author} {\bibfnamefont {J.}~\bibnamefont {Brivio}}, \bibinfo {author}
  {\bibfnamefont {M.}~\bibnamefont {Watson}}, \bibinfo {author} {\bibfnamefont
  {X.}~\bibnamefont {Wu}}, \bibinfo {author} {\bibfnamefont {A.}~\bibnamefont
  {Kis}}, \bibinfo {author} {\bibfnamefont {T.}~\bibnamefont {Luo}}, \bibinfo
  {author} {\bibfnamefont {A.~R.}\ \bibnamefont {{Hight Walker}}}, \ and\
  \bibinfo {author} {\bibfnamefont {H.~G.}\ \bibnamefont {Xing}},\ }\href@noop
  {} {\bibfield  {journal} {\bibinfo  {journal} {ACS Nano}\ }\textbf {\bibinfo
  {volume} {8}},\ \bibinfo {pages} {986} (\bibinfo {year} {2014})}\BibitemShut
  {NoStop}%
\end{thebibliography}%
%merlin.mbs apsrev4-1.bst 2010-07-25 4.21a (PWD, AO, DPC) hacked
%Control: key (0)
%Control: author (8) initials jnrlst
%Control: editor formatted (1) identically to author
%Control: production of article title (-1) disabled
%Control: page (0) single
%Control: year (1) truncated
%Control: production of eprint (0) enabled
%

%%%%%%%%%%%%%%%%%%%%%%%%%%%%%%%%%%%%%%%%%%%%%%%%%%%%%%%%%%%%%%%%%%%%%%%%%%%%%%%%%
\end{document}